\documentclass[draft]{agujournal2019}
\usepackage{url} %this package should fix any errors with URLs in refs.
\usepackage[inline]{trackchanges} %for better track changes. finalnew option will compile document with changes incorporated.
\usepackage{soul}
\journalname{Water Resources Research}

\usepackage{algorithm}
\usepackage{algorithmic}
\usepackage{tikz} 
\usepackage{amsmath,amssymb}
\usetikzlibrary{calc}

\usepackage{dsfont}
\usetikzlibrary{shapes.geometric}
\newcommand{\norm}[2][]{\left\Vert #2\right\Vert_{#1}}

\newcommand{\eps}{\boldsymbol{\varepsilon}}
\newcommand{\Eps}{\mathcal{E}}
\newcommand{\Nu}{\mathcal{V}}
\newcommand{\bs}[1]{\ensuremath{\boldsymbol{#1}}}

\newcommand{\kkf}{\bs{k_{\text{fine}}}}
\newcommand{\kkc}{\bs{k_{\text{coarse}}}}
\newcommand{\yp}{\bs{y_{\text{process}}}}

\newcommand{\yo}{\bs{y_{\text{obs}}}}
\newcommand{\f}{f}
\newcommand{\g}{g}

\newcommand{\kk}{\bs{k}}

\begin{document}

%% ------------------------------------------------------------------------ %%
%  Title
%
% (A title should be specific, informative, and brief. Use
% abbreviations only if they are defined in the abstract. Titles that
% start with general keywords then specific terms are optimized in
% searches)
%
%% ------------------------------------------------------------------------ %%

% Example: \title{This is a test title}

\title{Incorporating Posterior-Informed Approximation Errors into a Hierarchical Framework to Facilitate Out-of-the-Box MCMC Sampling for Geothermal Inverse Problems and Uncertainty Quantification}

%% ------------------------------------------------------------------------ %%
%
%  AUTHORS AND AFFILIATIONS
%
%% ------------------------------------------------------------------------ %%

% Authors are individuals who have significantly contributed to the
% research and preparation of the article. Group authors are allowed, if
% each author in the group is separately identified in an appendix.)

% List authors by first name or initial followed by last name and
% separated by commas. Use \affil{} to number affiliations, and
% \thanks{} for author notes.
% Additional author notes should be indicated with \thanks{} (for
% example, for current addresses).

% Example: \authors{A. B. Author\affil{1}\thanks{Current address, Antartica}, B. C. Author\affil{2,3}, and D. E.
% Author\affil{3,4}\thanks{Also funded by Monsanto.}}

%\authors{=list all authors here=}

\authors{Oliver J. Maclaren\affil{1}, Ruanui Nicholson\affil{1}, Elvar K. Bjarkason\affil{1}, John P. O'Sullivan\affil{1}\ \& Michael J. O'Sullivan\affil{1}}

 \affiliation{1}{Department of Engineering Science, The University of Auckland, Auckland, New Zealand}
% \affiliation{2}{Second Affiliation}
% \affiliation{3}{Third Affiliation}
% \affiliation{4}{Fourth Affiliation}

%\affiliation{=number=}{=Affiliation Address=}
%(repeat as many times as is necessary)

%% Corresponding Author:
% Corresponding author mailing address and e-mail address:

% (include name and email addresses of the corresponding author.  More
% than one corresponding author is allowed in this LaTeX file and for
% publication; but only one corresponding author is allowed in our
% editorial system.)

% Example: \correspondingauthor{First and Last Name}{email@address.edu}

\correspondingauthor{Oliver J. Maclaren}{oliver.maclaren@auckland.ac.nz}
\correspondingauthor{Ruanui Nicholson}{ruanui.nicholson@auckland.ac.nz}

%% Keypoints, final entry on title page.

%  List up to three key points (at least one is required)
%  Key Points summarize the main points and conclusions of the article
%  Each must be 100 characters or less with no special characters or punctuation 

% Example:
% \begin{keypoints}
% \item	List up to three key points (at least one is required)
% \item	Key Points summarize the main points and conclusions of the article
% \item	Each must be 100 characters or less with no special characters or punctuation
% \end{keypoints}

\begin{keypoints}

\item We consider geothermal inverse problems and uncertainty quantification from a Bayesian perspective.

\item We present a simple method for incorporating posterior-informed approximation errors into a hierarchical Bayesian framework.

\item Our method makes standard out-of-the-box MCMC sampling feasible for more complex models while correcting for bias and overconfidence.

\end{keypoints}

%% ------------------------------------------------------------------------ %%
%
%  ABSTRACT
%
% A good abstract will begin with a short description of the problem
% being addressed, briefly describe the new data or analyses, then
% briefly states the main conclusion(s) and how they are supported and
% uncertainties.
%% ------------------------------------------------------------------------ %%

%% \begin{abstract} starts the second page

\begin{abstract}
We consider geothermal inverse problems and uncertainty quantification from a Bayesian perspective. Our main goal is to make standard, `out-of-the-box' Markov chain Monte Carlo (MCMC) sampling more feasible for complex simulation models by using suitable approximations. To do this, we first show how to pose both the inverse and prediction problems in a hierarchical Bayesian framework. We then show how to incorporate so-called posterior-informed model approximation error into this hierarchical framework, using a modified form of the Bayesian approximation error (BAE) approach. This enables the use of a `coarse', approximate model in place of a finer, more expensive model, while accounting for the additional uncertainty and potential bias that this can introduce. Our method requires only simple probability modelling, a relatively small number of fine model simulations, and only modifies the target posterior -- any standard MCMC sampling algorithm can be used to sample the new posterior. These corrections can also be used in methods that are not based on MCMC sampling. We show that our approach can achieve significant computational speed-ups on two geothermal test problems. We also demonstrate the dangers of naively using coarse, approximate models in place of finer models, without accounting for the induced approximation errors. The naive approach tends to give overly confident and biased posteriors while incorporating BAE into our hierarchical framework corrects for this while maintaining computational efficiency and ease-of-use.  
\end{abstract}

%% ------------------------------------------------------------------------ %%
%
%  TEXT
%
%% ------------------------------------------------------------------------ %%

\section{Introduction}
%\subsection{Geothermal Modelling and Inverse Problems}
Computational modelling plays an important role in geothermal reservoir engineering and resource management. A significant task for decision making and prediction in geothermal resource management is so-called \textit{inverse modelling}, also known as \textit{model calibration} within the geothermal community, and as solving \textit{inverse problems} in applied mathematics. Calibration consists of determining parameters compatible with measured data. This is in contrast to so-called \textit{forward modelling} in which a simulation is based on known model parameters. Comprehensive reviews of geothermal modelling, including both forward modelling and model calibration, are given by \citeA{OSullivan2001,OSullivan2016}. 

The primary parameters of interest in geothermal inverse problems include the anisotropic permeability of the subsurface and the location and strength of so-called deep upflows/sources. Knowledge of the values of these parameters allows for forecasts of, for example the temperature and pressure down drilled, or to-be-drilled, wells, to be made. On the other hand, the available (i.e. directly measurable) quantities are instead typically temperature, pressure and enthalpy at observation wells \cite{OSullivan2001,OSullivan2016}. A typical geothermal inverse problem for a natural, i.e., steady, state, pre-exploitation model then consists of, for example, estimating formation permeabilities based on temperature and/or pressure measurements at observation wells. 

The predominant method used to solve geothermal inverse problems is still manual calibration \cite{OSullivan2016,Mannington2004,Burnell2012,OSullivan2009}, although it is well-recognised that this is far from an optimal strategy. To address this situation, there has been a concerted effort to automate the calibration process. For example, software packages such as iTOUGH2 \cite{Finsterle2000itough2} and PEST \cite{Doherty2015pest} have been developed, and used, for geothermal model calibration. These packages are primarily based on framing the inverse problem as one of finding the minimum of a regularised cost, or objective function; though essentially deterministic, approximate confidence (or credibility) intervals for model parameters can be constructed from local cost-function derivative information \cite{aster2018parameter}. Even for optimisation-based approaches to geothermal inverse problems, computations can be expensive and improvements are required to speed up the process. We recently proposed accelerating optimisation-based solution methods using adjoint methods and randomised linear algebra \cite{Bjarkason2017,Bjarkason2019pass,Bjarkason2019-xl}.

Bayesian inference is an alternative to optimisation-based approaches, and is instead an inherently probabilistic framework for inverse problems \cite{Tarantola2004,Kaipio2005,Stuart2010}. This naturally allows for incorporation and quantification of uncertainty in the estimated parameters; when posed in the Bayesian setting the solution to the inverse problem is an entire probability density over the parameters. Here we adopt a \textit{hierarchical} Bayesian approach in particular, where we use `hierarchical Bayes' in the sense of \citeA{Berliner1996hierarchical,Berliner2003physical,Berliner2012statistical}. This approach is discussed in detail in Section \ref{sec: Hierarchical}. The key to the method proposed here is incorporating approximation errors between an accurate and a coarse model as a component in our hierarchical framework, by adapting the Bayesian approximation error (BAE) approach \cite{Kaipio2005,Kaipio2013}. This allows us to speed up computation of parameter estimates, while avoiding overconfidence in biased estimates by accounting for the approximation errors induced when coarsened models are used. The trade-off for improved computation time is modified posteriors with inflated variance relative to the ideal target posterior.

There is only a relatively small amount of literature taking a fully Bayesian approach to geothermal inverse problems \cite<e.g.>{Cui2011,Maclaren2016,cui2019using,cui2019posteriori} where by `fully Bayesian' we mean sampling (or otherwise computing) a full probability distribution rather than calculating a single point estimate and making local approximations to the posterior covariance matrix. We previously presented a hierarchical Bayesian approach to frame the inverse problem, and used a generic sampling method to solve the resulting problem \cite{Maclaren2016}. On the other hand, \citeA{Cui2011} and \citeA{cui2019posteriori} developed a more sophisticated adaptive sampling scheme based on using a coarsened model and a fine model. The present work is based on extending the hierarchical Bayesian framework of \citeA{Maclaren2016} to explicitly use approximate models, while being independent of which sampling scheme is used and straightforward to implement.

\section{Background: The Bayesian Approach to Inverse Problems}
The Bayesian framework for inverse problems allows for systematic incorporation and subsequent quantification of parameter uncertainties \cite{Kaipio2005,Stuart2010}, which can then be propagated through to model predictions. In this framework, the solution to the inverse problem is an entire probability distribution, i.e., the {\em posterior probability distribution}, or simply {\em the posterior}. Both epistemic (knowledge-based) and aleatoric (actually random) uncertainties are represented using the same probabilistic formalism in Bayesian inference.

Calculation of the posterior relies on Bayes' theorem, written here as 
\begin{equation}
p(\kk\vert \yo) \propto p(\yo \vert \kk)p(\kk),
\label{eq: BayesThm}
\end{equation}
where $\kk$ denotes the parameters of interest and $\yo$ denotes measured data, such as downhole temperatures. Here our parameters of interest here are rock permeabilities; though we work with log permeabilities throughout, for simplicity we will generally refer to these simply as `permeabilities'. The above is written as a proportionality relationship leaving out a normalisation factor that is not required for most sampling algorithms \cite{gelman2013bayesian}. In the above, $p(\yo | \kk)$ is termed the \textit{likelihood} and $p(\kk)$ is the \textit{prior}.

A drawback of the fully Bayesian approach is the intensive computational cost that is usually required to apply Bayes' theorem, especially in the case of complex models such as in the geothermal setting \cite<see for example>{cui2015data}. The dominant cost is repeated evaluation of the forward model and thus coarsened or surrogate models are often used in place of the most accurate forward model \cite<see for example>{asher2015review}. Furthermore, the use of coarsened or surrogate models can help alleviate numerical instabilities \cite{Doherty2011-zw}. However, replacement of an accurate model with a surrogate invariably results in so-called {\em approximation errors}, which, if not accounted for, can lead to parameters and their associated uncertainty being incorrectly estimated \cite<see for example>{Kennedy2000-di,Kaipio2007,doherty2010short}. Next we give a brief overview of the main approaches in the literature for accounting for these errors. We then discuss how we incorporate these ideas into a hierarchical framework.

\subsection{Approximation Errors and Model Discrepancies}
In the Bayesian viewpoint, approximation errors can be treated as a further source of uncertainty. There are two standard approaches for doing this, i.e. dealing with approximation errors: that based on the work of \citeA{Kennedy2000-di} (referred to as KOH hereafter), and the BAE approach proposed by \citeA{Kaipio2005}. The underlying principles of both approaches are similar, though with some implementation and philosophical differences. In particular, the KOH method was explicitly developed to account both for the difference between `reality' and a given simulation model, and to allow for efficient emulation of computationally expensive models at arbitrary values \cite{Kennedy2000-di,higdon2004combining,higdon2008computer}. The typical KOH method is based on infinite-dimensional Gaussian process (GP) models; one to model the difference between reality and the simulation model, and one to represent the output of the simulation model at new input values. Usually only one physically-based model is used \cite{higdon2004combining,higdon2008computer}. 

The BAE approach, in contrast, is based on two physically-based simulation models: one which represents the `best', but typically very expensive model, and one representing a coarser model, which nevertheless preserves the key physics of the problem. Furthermore, the approximation errors between the two physically-based models are represented by a finite-dimensional multivariate Gaussian distribution, defined only at the locations of interest. The statistics of the approximation errors are directly estimated empirically, based on a small number of simulations of both the accurate and coarse models, and structural constraints are not typically placed on the form of the covariance matrix \cite{Kaipio2013}. While differences between the fine model and coarse model in the BAE approach are generally considered `approximation' errors between two different models, rather than `discrepancies' between reality and a model as in the KOH approach, these approximation errors typically include significant correlation structure, and the approach has been shown to work well in physical experiments \cite<see for example>{Nissinen2008,lipponen2011nonstationary,nissinen2010compensation}. Additional systematic error can also be incorporated in the BAE approach in a straightforward manner, i.e. it can directly incorporate correlation structure for both the error between the fine model and the data, and in the error between the fine model and the coarse model. 

%these approximation errors can be considered either as proxies for the sort of discrepancies one might expect when using an inadequate model, or at least as compensating to some extent for these unmodelled errors. In particular, 

The BAE approach is particularly simple to implement and, given two physical models, requires less user input in terms of parameters and hyperparameters than the KOH approach. For a further discussion and comparison of the the two methods see \citeA{ColinFoxPaper}. In this work we use (a variant of) the BAE approach; in contrast to past work in this area, however, we explicitly incorporate the approximation errors into a hierarchical framework. We discuss this next.

\section{Hierarchical Framework}\label{sec: Hierarchical}
Here we outline our hierarchical Bayesian framework and where approximation errors enter. Implementation details are given in the following sections.

As described in \citeA{Maclaren2016}, the hierarchical Bayesian approach generally begins by assuming a three-stage decomposition of a full joint probability distribution over all quantities of interest, written schematically as;
\begin{equation}
\begin{aligned}
&p(\text{observed data, } \text{process variable, process parameters, observation parameters})\\
&= \\
&p(\text{observed data } \vert \text{ process variable, observation parameters})\\
&\times\\
&p(\text{process variable } \vert \text{ process parameters})\\
&\times \\
&p(\text{process parameters, observation parameters}).
\end{aligned}
\end{equation}

\noindent \\These three stages correspond to a measurement model, process model, and parameter model, respectively. The process parameters and observation parameters include both parameters of interest, such as permeabilities, and parameters characterising the probability distributions, such as covariance matrices. The above decomposition is not an identity of probability theory, but instead contains plausible physical \textit{modelling} assumptions about the conditional independencies separating measurement and process variables \cite{Berliner1996hierarchical,Berliner2003physical,Berliner2012statistical}. For example, the measurement model (first factor) is assumed to be independent of the process parameters, while the process model (second factor) is assumed to be independent of the observation parameters. In terms of our current problem variables this becomes
\begin{equation}
p(\yo, \yp, \kk) = p(\yo \vert \yp)p(\yp \vert \kk)p(\kk),
\label{eq: hierarchical_fact}
\end{equation}
where $\yo$ is the observable (hence noisy) data vector, $\yp$ is the latent or `true' process vector, and we have suppressed the distribution parameters and distribution subscripts in each stage for simplicity. The model approximation error enters into the above scheme as a probabilistic process error. Intuitively, we use a probabilistic model to capture the additional uncertainty introduced by using an approximate model in place of a more accurate model. This is despite the fact that both models are deterministic; we explain the nature of this approximation in the following sections. Note that in order to regain the standard form of Bayes' theorem, i.e. Equation (\ref{eq: BayesThm}), the process variable $\yp$ can be marginalised (integrated) out to regain the likelihood, $p(\yo \vert \kk)$. This procedure is described in detail below.

\subsection{Representation Using Functional Relationships}
An equivalent representation of the above factorisation scheme can be given in terms of functional relationships between random variables. In particular, assuming additive error models, the measurement and process model components correspond to a two-stage decomposition of the form
\begin{equation}
    \begin{aligned}
    &\yo =\ \yp + \bs{e}\\
    &\yp =\ \g(\kk) + \bs{\eps}.
    \end{aligned}
    \label{eq: twostage_func}
\end{equation}
where $\bs{e}$ is the measurement error (which may include correlations), and $\g$ represents an approximate process model, i.e. one that introduces additional errors $\bs{\eps}$, which as alluded to previously still preserves the key physics. Again these may include significant correlations, though we will assume that the two vectors $\bs{e}$ and $\bs{\eps}$ are independent of each other. 

Combining the error models, and introducing the {\em total error} term $\bs{\nu} = \bs{\eps} + \bs{e}$, allows us to write the above two-level model as a single-level model 

\begin{equation}
\yo = \g(\kk) + \bs{\eps} + \bs{e} = \g(\kk) + \bs{\nu},
\label{eq: totalerror_func}
\end{equation}
where the total error typically has non-trivial correlation structure due to approximation errors, and possibly also due to measurement errors. 

\subsection{Likelihood}
The above relationships define our likelihood $p(\yo | \kk)$, with the probability model dependent on the probabilistic structure of the total error vector $\bs{\nu}$. This likelihood can be obtained conveniently by using the single-stage functional relationship in Equation \ref{eq: totalerror_func} involving the total error, along with formal marginalisation over the total error: 

\begin{equation}
\begin{aligned}
p_{Y_o | K}(\yo | \kk) &= \int p_{Y_o, \Nu | K}(\yo, \bs{\nu} | \kk) d\bs{\nu} = \int p_{Y_o | \Nu, K}(\yo | \kk, \bs{\nu})p_{\Nu | K}(\bs{\nu} | \kk) d\bs{\nu}\\
&= \int \delta(\yo - g(\kk) - \bs{\nu})p_{\Nu | K}(\bs{\nu} | \kk) d\bs{\nu}\\
&= p_{\Nu | K}(\yo - g(\kk) | \kk)\\
&= p_{\Nu }(\yo - g(\kk)).
\end{aligned}
\end{equation}
The last step above follows when both error vectors are independent of the parameter, and $\delta$ is used to denote the Dirac delta distribution, which places all mass at 0. The assumption of independence of the model error vector and the parameter vector is discussed in detail in Section \ref{sec: error}. We have also explicitly denoted which probability distribution is being evaluated using subscripts.

These steps can be considered as a change of variables from $Y_o | K$ to $\Nu | K$ via the delta method \cite{Au1999,Khuri2004}. As would be expected, this likelihood can also be obtained by marginalising out the process variable in the factorisation given in Equation \ref{eq: hierarchical_fact}, and using the equivalent two-stage representation of the hierarchical model in Equations \ref{eq: twostage_func}, but the above derivation is slightly simpler.

\subsection{Error Components in the Hierarchical Framework}
Here we consider the two key sources of error, measurement error and process (approximation) error, in more detail.

\subsubsection{Measurement Error}
Our measurement model is assumed to be independent of the approximation errors, and takes the form
\begin{equation}
p_{Y_{o}|Y_{p}}(\yo | \yp) = p_{E|Y_{p}}(\yo - \yp | \yp) = p_{E}(\yo - \yp) = p_E(\bs{e}).
\end{equation}
Here
\begin{equation}
\bs{e} = \yo - \yp,
\end{equation}
is the measurement error, where $\bs{e} \sim p_{E | Y_{p}}(\cdot) = p_{E}(\cdot)$ due to our independence assumptions between the measurement and process error vectors. 

In the two physically motivated cases considered in this paper we make the assumption that the measurement errors are also pairwise independent. However, this assumption is not required, and changing it simply changes the covariance matrix of the measurement errors (assuming they are Gaussian). A simple example using correlated measurement noise is provided in \ref{sec: A3}.

\subsubsection{Process Error}
Process errors, i.e. approximation errors, are introduced by using a coarse model in place of a finer, or more accurate, simulation model. The fine model is represented by a function $\f(\kkf)$, and the coarse model by a function $\g(\kkc)$. Here $\kkf$ and $\kkc$ are the (vectors of) fine- and coarse-scale parameters of interest; in our case the fine-scale and coarse-scale parameters will have the same dimension here, despite corresponding to different discretisation grids. That is, both models share the same parameters, and thus we will drop the explicit distinction between $\kkf$ and $\kkc$ in what follows and simply refer to both by $\kk$ (but see \ref{sec: A1} for a discussion of the relationship between fine-scale and coarse-scale parameter grids). 

To model the process approximation error, we assume the true or latent process variable to be generated exactly by the fine-scale model, i.e.
\begin{equation}
\yp = \f(\kk).
\end{equation}
In \citeA{Maclaren2016} we only used one model and essentially had $p(\yp | \kk) = \delta(\yp - \g(\kk))$ as our process model. Here we explicitly introduce both fine and coarse models and take into account that $p(\yp | \kk) \neq \delta(\yp - \g(\kk))$. To do this, we define the process model error variable by 
\begin{equation}
\eps = \yp - \g(\kk) = \f(\kk) - \g(\kk).
\end{equation}
Since both $\f$ and $\g$ are deterministic for a given $\kk$, at this point $\eps$ must be too; this can be formally incorporated into the hierarchical model by again treating deterministic functions as delta distributions. Thus we can write
\begin{equation}
p_{Y_{p} | K}(\yp | \kk) = p_{\Eps | K}(\yp - \g(\kk) | \kk) = p_{\Eps | K}(\f(\kk) - \g(\kk) | \kk) = p_{\Eps | K}(\eps | \kk),
\end{equation}
which simply amounts to a deterministic change of variables (carried out, e.g., via the delta method \cite{Au1999,Khuri2004}). 

\subsubsection{Total Error}
Our goal here is to compute the posterior for the parameters given the data
\begin{equation}
p(\kk | \yo) \propto p(\yo | \kk)p(\kk),
\end{equation}
where the total error has been marginalised over, and where $p(\yo | \kk)$ is the likelihood and $p(\kk)$ is the prior. From the above, we see the posterior can be written as
\begin{equation}
\begin{aligned}
p_{K | Y_o}(\kk\ |\ \yo) &\propto p_{Y_o | K}(\yo | \kk) p_K(\kk) \\ 
&= p_{\Nu}(\yo - g(\kk))p_K(\kk),
\end{aligned}
\end{equation}
where the process error has now been absorbed into the likelihood. The resulting expression is hence simply a standard `measurement' likelihood function, written in terms of total error, multiplied by the prior. It does, however, require the distribution of the total error, $p_{\Nu}(\cdot)$, to be known. 

%&= p_{\Nu | K}(\yo - g(\kk) | \kk)p_K(\kk)\\

To construct a model of the total error, we a) assume that the measurement error $\bs{e}$ is Gaussian and b) approximate the process model error $\bs{\eps}$ as Gaussian. Both of these random vectors may in general exhibit significant correlations between their respective components, and this is accounted for in the present approach, but the two vectors are assumed independent of each other. That is, we assume process error and measurement error are independent of each other. This makes combining these two errors straightforward (as described in the next subsection). Ultimately, we determine whether these, and the other approximations used thus far, are reasonable based on whether they work in practice -- e.g. whether they recover good estimates of the true parameters in test cases, and whether any available error distributions `look normal' when plotted (or, if desired, pass formal tests of normality). 

\section{Computation of Error Models: Standard, Composite and Posterior-Informed}\label{sec: error}
Both the probabilistic process model $p_{Y_{p} | K}(\yp | \kk)$ and process error model $p_{\Eps | K}(\eps | \kk)$ are typically intractable to simulate from for more than a limited number of realisations, as both involve the expensive fine-scale model. This motivates using approximations to these distributions, and results in approximate posterior distributions relative to the ideal target. The goal of these approximations is not to accurately estimate the model error as such, but to approximately model the effect of marginalising over it. This is for the purpose of reducing the bias/overconfidence in parameter estimates that would result from just using the simpler model directly; some loss of precision/statistical efficiency is expected. Here we give an overview of how the standard, composite, and posterior-informed approximation error models are computed, and discuss relevant related literature. Explicit algorithms are given in the following section.

\subsection{Pre-marginalisation} 
Due to the computational issues discussed above, in the standard BAE approach the statistics of the approximation errors are pre-computed empirically via directly drawing samples from the prior distribution, without the use of MCMC. Similarly, here we compute the statistics of the approximation errors via direct sampling, though from a (naive) posterior distribution rather than the prior distribution, which itself can be (and, here, was) computed by separate MCMC sampling. MCMC sampling methods are discussed in Section \ref{sect: MCMC Sampling}. 

Our approach has the advantage of allowing a set budget of fine model runs to be specified, as well as minimal implementation difficulty. In contrast, some recent MCMC sampling schemes explicitly estimate and incorporate approximation errors \textit{during} the MCMC sampling process. Similarly to our proposed method, \citeA{Cui2011} and \citeA{cui2019posteriori} consider carrying out MCMC sampling on models of distinct levels of discretisation while accounting for the approximation error; however, they use an adaptive delayed acceptance (ADA) BAE approach to build the approximation error model during the MCMC sampling. In the methods developed by \citeA{Cui2011} and \citeA{cui2019posteriori} the accurate model is typically run for each MCMC sample accepted based on the coarse model. This can make it more difficult to control the number of fine model runs used. While it is possible in principle to further modify the MCMC scheme used to incorporate such constraints, our approach offers a simple and direct way of controlling the number of fine model runs used.

\citeA{xu2017quantifying}, \citeA{zhang2018inverse}, and \citeA{lodoen2010bayesian} apply the KOH method to account for the approximation errors, and these are incorporated into an {\em adaptive multifidelity} MCMC sampler \cite<see for example>{peherstorfer2018survey}, the {\em differential evolution adaptive metropolis (DREAM)} \cite{vrugt2009accelerating,laloy2012high} sampler, and the Metropolis-Hastings (MH) algorithm \cite<see for example>{chib1995understanding}, respectively. Again, these require more sophisticated understanding and control of the MCMC scheme used, and involve infinte-dimensional stochastic processes following the approach of KOH. Here we provide a simple alternative based on the BAE approach to approximation error and involving finite-dimensional probability distributions only.

\subsection{Standard Approximation Error}
The standard BAE approach \cite{Kaipio2005,Kaipio2013} is to first simulate a limited number of realisations from the \textit{true} (i.e. involving the fine-scale model) joint distribution
\begin{equation}
p_{\Eps, K}(\eps,\kk) = p_{\Eps | K}(\bs{\eps} | \kk)p_{K}(\kk) 
\end{equation}
using a given parameter prior $p_{K}(\kk)$, and then fit an approximate distribution $\hat{p}_{\Eps, K}(\bs{\eps},\kk)$ to the $(\bs{\eps}, \kk)$ realisations. This empirically-estimated approximate distribution is then used as a plug-in replacement
\begin{equation}
    p_{\Eps, K}(\bs{\eps},\kk)  \leftarrow \hat{p}_{\Eps, K}(\bs{\eps},\kk)
\end{equation}
in the hierarchical model. While the true points will lie on a surface of zero thickness, $\hat{p}$ is estimated within a non-degenerate family of probability distributions, such as a multidimensional normal distribution. This procedure aims to `conservatively' cover the sample points, despite the obvious model mis-specification (see Figure \ref{fig:composite_error}). 

\subsection{Enhanced, or Composite, Approximation Error}
A further approximation is often used, which leads to what is called the `enhanced error model' in the BAE literature \cite{Kaipio2005,Kaipio2013}, which we will follow in the present work. This amounts to replacing the true joint distribution by the product of the empirically-estimated -- but true -- marginal distributions:
\begin{equation}
    p_{\Eps, K}(\eps,\kk)  \leftarrow p_{\Eps}(\bs{\eps})p_{K}(\kk)
\end{equation}  
where
\begin{equation}
    p_{\Eps}(\bs{\eps}) =  \int p_{\Eps | K}(\bs{\eps} | \kk)p_{K}(\kk) d\kk,
\end{equation}
which is estimated empirically based on samples as described in Section \ref{sec: ErrStats}. In the above, estimation of $p_{\Eps}(\bs{\eps})$, $p_{\Eps | K}(\bs{\eps} | \kk)$ is taken as the \textit{true} conditional error distribution, and hence the samples are used to estimate the \textit{true} marginal. On the other hand, in all subsequent calculations the joint distribution is approximated by the product of the marginals. This is equivalent to using the marginal error distribution for $\eps$ as a plug-in empirical estimator of the conditional error distribution for $\bs{\eps} | \kk$ in the hierarchical model, prior to subsequent inference steps. Importantly, this does not mean that the individual errors in the vector $\eps$ are independent of each other, rather that the vector random variable $\bs{\eps}$ is independent of the vector random variable $\kk$. The estimated errors $\bs{\eps}$ almost always exhibit significant correlations between components, and these are accounted for here.

As emphasised above, the goal is not to get the error exact, but to account for it in a somewhat `conservative' manner. While in the BAE literature this is referred to as the enhanced error model, the replacement of an intractable conditional distribution in a product of distributions by a more accessible marginal distribution is also similar in philosophy to that used in, for example, the composite likelihood literature \cite{Varin2008,Varin2011}. Hence we will prefer to refer to it as the \textit{composite} error model in the remainder of the text. 

Finally, we note that \textit{after} both the true marginal process model error has been empirically estimated, and the plug-in replacement has been made for the conditional distribution, the full process model error vector, $\bs{\eps}$, is assumed to be (formally) conditionally independent of the full parameter vector, $\kk$, in any subsequent manipulations of the probability distributions. 

\subsection{Posterior-Informed Composite Approximation Error}
Another practical issue with both of the above approximation procedures (i.e. both the full and the composite error models) arises in complicated models such as those in geothermal reservoir modelling \cite<see e.g.>{OSullivan2016}: model run failures, long model run times and/or extreme model outputs when sampling from an insufficiently informative prior and running the fine-scale model (in particular). We encountered a large number of such model run issues for the fine-scale model and were thus motivated to consider a further approximation to the process model error. This can be described as a \textit{posterior plug-in} estimate of the model approximation error. In particular, we make the plug-in estimate  
\begin{equation}
    p_{\Eps}(\bs{\eps})  \leftarrow \hat{p}_{\Eps|Y_o}(\bs{\eps} | \yo),
\end{equation}
where we now use the coarse-model posterior for the parameters to estimate the error distribution marginalised over the parameter. That is, we use
\begin{equation}
    \hat{p}_{\Eps|Y_o}(\eps | \yo) =  \int p_{\Eps | K}(\eps | \kk)\hat{p}_{K | Y_o}(\kk | \yo) d\kk 
\end{equation}
which is estimated empirically based on samples as described in Section \ref{sec: ErrStats}, and where
\begin{equation}
\hat{p}_{K | Y_o}(\kk | \yo) \propto \hat{p}_{Y_o | K}(\yo | \kk)p(\kk)
\end{equation}
and $\hat{p}_{Y_o | K}$ is the likelihood function based on the coarse-scale model $g(\kk)$. Since we did not encounter model run issues in the coarse model we can estimate this by combining the likelihood with the broad prior.

Once the error distribution has been estimated we again use the composite model of the joint distribution, along with the original prior:
\begin{equation}
p_{\Eps, K}(\eps,\kk)  \leftarrow p_{\Eps}(\eps)p_{K}(\kk).
\end{equation}
Thus we are simply using a different plug-in estimate of the model error. Since this is now used with the coarse model we can revert to the broad prior without model run failures in all subsequent calculations.

Again, because our goal is not to model the error exactly, but rather to model the effect of marginalising over it, we are willing to tolerate more potential inaccuracies at this stage. The present step of using posterior sampling for the approximation error is `riskier' than that in the previous section, however, in the sense that it involves a formal `double use of data' and tends to \textit{narrow} rather than widen the error distribution, when compared to the distribution that results from using the prior. A geometric interpretation of this posterior model approximation step, and its potential dangers, is given in \ref{sec: A2}. 

Despite the above warnings we believe that the use of posterior approximation errors, as described in the present work, is often a practical solution in complex models. It also has the benefit of providing more `relevant' estimates of the model error when the posterior based on the coarse model is not too far from the true posterior. One way to check this assumption would be to recompute the model error distribution under the final posterior and compare it to the error distribution computed under the coarse model posterior; checking for similarity of these distributions can be thought of as a form of posterior predictive check \cite<see e.g.>[for a good general discussion of posterior predictive checks]{gelman2013bayesian}. This check does, however, require recomputing realisations from the fine-scale model and so is not always practical.

\section{Statistical Algorithms}\label{sec: ErrStats} 
By taking a Gaussian approximation of the process error we can characterise its distribution with the mean and covariance only. As discussed above, these cannot be computed analytically in general, and thus must be estimated empirically via samples. In this section we give algorithmic details for both the standard  {\em composite error model}  approach and our proposed {\em posterior-informed composite error model} approach. Pseudocode is provided for both of the methods. We also outline the MCMC method used for sampling the resulting target posterior.

\subsection{The Standard Composite Error Model Approach}
To calculate the statistics of the the process error, $\bs{\eps}$, in the standard composite error model approach, an ensemble of $q\in\mathbb{N}$ samples are drawn from the prior distribution $p(\kk)$, say $\kk^{(\ell)}$, for $\ell=1,2,\dots,q$. Both the fine and coarse models are then run for these samples, resulting in an ensemble of approximation errors
\begin{align}\label{eq: disc}
\bs{\eps}^{(\ell)}=\f(\kk^{(\ell)})-\g(\kk^{(\ell)}).
\end{align}
The ensemble mean and covariance of the approximation errors are then estimated,
\begin{align}\label{eq: ensemblestats}
\bs{\eps}_*=\frac{1}{q}\sum_{\ell=1}^q \bs{\eps}^{(\ell)},\quad \bs{\Gamma}_{\eps}=\frac{1}{q-1}\sum_{\ell=1}^q (\eps^{(\ell)}-\bs{\eps}_*)(\bs{\eps}^{(\ell)}-\bs{\eps}_*)^T.
\end{align}
As discussed above, the total error, $\bs{\nu}$, is the sum of both the noise and the process model error, and thus the distribution for the total error (given the normality assumption) is given by
\begin{align}\label{eq: totalerrordist}
\bs{\nu}\sim\mathcal{N}(\bs{\nu}_*,\bs{\Gamma}_\nu)=\mathcal{N}(\bs{e}_*+\bs{\eps}_*,\bs{\Gamma}_e+\bs{\Gamma}_{\eps}).
\end{align}
This new distribution is then used to update the likelihood, which consequently updates the posterior density.

Algorithm \ref{alg: 1} gives pseudocode for the standard composite error model approach for constructing the distribution of the total errors and for carrying out the inversion.

\begin{minipage}{0.9\linewidth}
\begin{center}
\begin{algorithm}[H]
\caption{Standard {\it composite error model} approach}\label{alg: 1}
\begin{algorithmic} 
\STATE Generate $\kk^{(1)},\kk^{(2)},\dots,\kk^{(q)}$ from the prior, $p(\kk)$
\STATE Set $\bs{\eps}^{(\ell)}=f(\kk^{(\ell)})-g(\kk^{(\ell)})$ for $\ell=1,2,\dots,q$
\STATE Calculate $\bs{\eps}_*=\frac{1}{q}\sum_{\ell=1}^q \bs{\eps}^{(\ell)}$ and  $\bs{\Gamma}_{\eps}=\frac{1}{q-1}\sum_{\ell=1}^q (\bs{\eps}^{(\ell)}-\bs{\eps}_*)(\bs{\eps}^{(\ell)}-\bs{\eps}_*)^T$
\STATE Substitue $\bs{\nu}=\bs{\eps}+\bs{e}$ so that $\bs{\nu}\sim\mathcal{N}(\bs{\nu}_*,\bs{\Gamma}_\nu)=\mathcal{N}(\bs{\eps}_*+\bs{e}_*,\bs{\Gamma}_{\eps}+\bs{\Gamma}_e)$ 
\STATE Replace likelihood $p_E(\yo-f(\kk))$ with $p_\Nu(\yo-g(\kk))$
\STATE Compute revised posterior, $\propto p_\Nu(\yo-g(\kk))p(\kk)$, based on updated likelihood 
\end{algorithmic}
\end{algorithm}
\end{center}
\end{minipage}

\subsection{The Proposed Posterior-Informed Composite Error Model Approach}
In the approach proposed here, we avoid sampling from the prior density of $\kk$ to generate the ensemble $\eps^{(\ell)}$, to avoid model failures and extreme run times. Instead, we initially construct a {\em naive posterior} density of $\kk$, denoted $\hat{p}(\kk|\yo)$, (done here) using MCMC with the likelihood function induced by the noise term, $\bs{e}$, only, and using the coarse model, $g$. This results in samples from the naive posterior, $\kk^{(\ell)}$, which are then passed through the two models to construct the process model errors, $\eps^{(\ell)}$. Once these samples for $\eps$ have been generated, the method is essentially the same as that of the standard composite error model approach.

Pseudocode for the proposed posterior-informed composite error model approach is given in Algorithm \ref{alg: 2}.

\begin{minipage}{0.9\linewidth}
\begin{center}
\begin{algorithm}[H]
\caption{Proposed {\it posterior-informed composite error model} approach}\label{alg: 2}
\begin{algorithmic} 
\STATE Generate $\kk^{(1)},\dots,\kk^{(q)}$ from naive posterior $\hat{p}(\kk|\yo)$, using coarse model $g$
\STATE Set $\bs{\eps}^{(\ell)}=f(\kk^{(\ell)})-g(\kk^{(\ell)})$ for $\ell=1,2,\dots,q$
\STATE Calculate $\eps_*=\frac{1}{q}\sum_{\ell=1}^q \eps^{(\ell)}$ and  $\bs{\Gamma}_{\eps}=\frac{1}{q-1}\sum_{\ell=1}^q (\bs{\eps}^{(\ell)}-\eps_*)(\eps^{(\ell)}-\eps_*)^T$
\STATE Substitute $\bs{\nu}=\bs{\eps}+\bs{e}$ so that $\bs{\nu}\sim\mathcal{N}(\bs{\nu}_*,\bs{\Gamma}_\nu)=\mathcal{N}(\bs{\eps}_*+\bs{e}_*,\bs{\Gamma}_{\eps}+\bs{\Gamma}_e)$ 
\STATE Update likelihood, from $p_E(\yo-f(\kk))$ to $p_\Nu(\yo-g(\kk))$
\STATE Compute revised posterior, $\propto p_\Nu(\yo-g(\kk))p(\kk)$, based on updated likelihood 
\end{algorithmic}
\end{algorithm}
\end{center}
\end{minipage}

%This is also apparent from the replacement of a marginal distribution by a conditional distribution (conditional on the observations), rather than the usual replacement of a conditional distribution by a marginal distribution. We are still ultimately replacing a deterministic function by a probabilistic model, however.

\subsection{MCMC Sampling}\label{sect: MCMC Sampling}
In the present work, MCMC sampling is carried out using the Python package {\it emcee} \cite{Foreman-Mackey2013}. This package implements an affine invariant ensemble sampler \cite{Goodman2010}, with the benefit of being easy to implement for arbitrary user-defined models. It also allows for easy communication with the PyTOUGH Python interface \cite{Croucher2011} to TOUGH2 \cite{Pruess1999} and AUTOUGH2 \cite{Yeh2012} (The University of Auckland's own version of the TOUGH2) for carrying out the forward simulations.  

For large dimensional problems the affine invariant ensemble sampler may be inadequate \cite{Huijser2015}, in which case, alternative out-of-the-box samplers like those available in Stan \cite{carpenter2017stan}, or PyMC \cite{patil2010pymc} could be used. However, as alluded to earlier, the approach outlined here is essentially independent of the choice of particular MCMC sampler, providing flexibility in the choice of MCMC sampling scheme used while also being compatible with non-sampling, optimization-based methods.

\section{Computational Studies}
We consider multiphase nonisothermal flow in a geothermal reservoir, including both two-dimensional and three-dimensional reservoir case studies. 

\subsection{Governing Equations for Geothermal Simulations}
Our general problem is governed by the mass balance and the energy balance equations:
\begin{align}\label{eq: forward1}
\frac{{\rm d}}{{\rm d}t}\int_\Omega M_{\rm m}\;dV= -\int_{\partial\Omega}\bs{F}_{\rm m}\cdot\bs{n}\;dS+\int_\Omega q_{\rm m}\;dV
\end{align}
and
\begin{align}\label{eq: forward2}
\frac{{\rm d}}{{\rm d}t}\int_\Omega M_{\rm u}\;dV= -\int_{\partial\Omega}\bs{F}_{\rm u}\cdot\bs{n}\;dS+\int_\Omega q_{\rm u}\;dV,
\end{align}
respectively \cite{Pruess1999}. Here $\Omega$ is the control volume with boundary $\partial \Omega$, $\bs{n}$ denotes an outward pointing unit normal vector to $\partial\Omega$, $M_{\rm m}$ and $M_{\rm u}$ represent amount of mass per unit volume (kg/m$^3$) and amount of energy per unit volume (J/m$^3$) respectively, $\bs{F}_{\rm m}$ and $\bs{F}_{\rm u}$ are the mass flux (kg/(m$^2\cdot$s)) and energy flux respectively (J/(m$^2\cdot$s)), while $q_{\rm m}$  and $q_{\rm u}$ represent mass sinks/sources (kg/(m$^3\cdot$s)) and energy sinks/sources (J/(m$^3\cdot$s)) respectively. 

We consider observations of temperature only, while the parameters of interest are limited to (log) rock type permeabilities. Other parameters which may be of interest include deep sources, relative permeabilities, and porosities, while other observable quantities include production history pressure and enthalpy. The relationship between the permeabilities and temperature, i.e. the {\em parameter-to-observable map}, can be understood by examining the key terms in (\ref{eq: forward1}) and (\ref{eq: forward2}), following \citeA{Cui2011}. A more in-depth discussion is given in \cite{Pruess1999}. Firstly, the amount of mass and energy per unit control volume are given by 
\begin{align}
M_{\rm m}&=\phi(\rho_{\rm l}S_{\rm l}+\rho_{\rm v}S_{\rm v}),\\
M_{\rm u}&=(1-\phi)\rho_{\rm r}u_{\rm r}T+\phi(\rho_{\rm l}u_{\rm l}S_{\rm l}+\rho_{\rm v}u_{\rm v}S_{\rm v}),
\end{align}
respectively, where $\phi$ is porosity (dimensionless), $S_{\rm l}$ denotes liquid saturation  (dimensionless), $S_{\rm v}$ represents vapour saturation  (dimensionless), $\rho_{\rm l}$ signifies the density of the liquid (kg/m$^3$), $\rho_{\rm v}$ is the vapour density (kg/m$^3$), $\rho_{\rm r}$ represents density of the rock (kg/m$^3$), $u_{\rm l}$ denotes internal energy of the liquid, $u_{\rm v}$ signifies internal energy of the vapour (J/kg), $u_{\rm r}$ is the specific heat of the rock (J/kg$\cdot$K), and $T$ denotes temperature (K). Next, the mass flux is given by the sum of the mass flux of liquid and the mass flux of vapour,
\begin{align}
\bs{F}_{\rm m}=\bs{F}_{\rm ml}+\bs{F}_{\rm mv},
\end{align}
where 
\begin{align}
\bs{F}_{\rm ml}&=-\frac{\mathbf{K}k_{\rm rl}}{\nu_{\rm l}}(\nabla p-\rho_{\rm l}\bs{g}),\\
\bs{F}_{\rm mv}&=-\frac{\mathbf{K}k_{\rm rv}}{\nu_{\rm v}}(\nabla p-\rho_{\rm v}\bs{g}).
\end{align}
Here $\mathbf{K}$ represents the permeability tensor (m$^2$), $p$ is pressure (Pa), $\nu_{\rm l}$ and $\nu_{\rm v}$ are kinematic viscosity of liquid (m$^2$/s) and vapour (m$^2$/s) respectively, $k_{\rm rl}$ and $k_{\rm rv}$ signify relative permeabilities (dimensionless), while $\bs{g}$ denotes gravitational acceleration (m/s$^2$). Finally, the energy flux is given by 
\begin{align}
\bs{F}_{\rm u}=h_{\rm l}\bs{F}_{\rm ml}+h_{\rm v}\bs{F}_{\rm mv}-K\nabla T,
\end{align}
where $h_{\rm l}$ and $h_{\rm v}$ are specific enthalpies (J/kg) of liquid and vapour, respectively, and $K$ represents thermal conductivity (J/(K$\cdot$m$\cdot$s)). In our study, all parameters other than rock permeabilities were taken as known.

\subsection{Model Setup and Simulation}
We consider two scenarios as case studies -- the first is based on a synthetic two-dimensional slice model, while the second is based on the Kerinci geothermal system, Sumatra, Indonesia. Each case study involves both a fine model and a coarse model, and thus, in total, we have four computational geothermal models in this work.

In all cases we solve the forward problem using the computer package AUTOUGH2 \cite{Yeh2012}, The University of Auckland's version of the TOUGH2 \cite{Pruess1999} simulator, with the pure water equation of state model, i.e., EOS1. We only consider steady state conditions, though, as standard, we calculate steady states via time marching to assist convergence to proper model solutions. 

The parameters of interest in both case studies are rock permeabilities, which are associated to a given rock type. There has been some work on allowing a distinct rock type for each cell in the computational model \cite{Cui2011,cui2019posteriori,Bjarkason2017,Bjarkason2019-xl}. However the standard approach in geothermal modelling and inversion \cite{OSullivan2016,popineauintegrated,witter2018value,fullagar2007towards}, and the approach taken here, is to base the simulation model on a conceptual model of the geological structure. The simulation hence respects the lithologic boundaries of these geological models. Mathematically this is equivalent to {\em regularisation by discretisation}, see for example \citeA{Kaipio2005} or \citeA{aster2018parameter}, and is a way of incorporating important prior information. The present approach can allow for arbitrary assumptions on the rock type structure, though at the cost of higher dimensionality and/or increased ill-posedness. We aim to investigate the effects of including uncertainty in geological structure in future studies.

\subsubsection{Case Study I: Slice Model}
\begin{figure}[H]
\centering
\includegraphics[width=.8\textwidth]{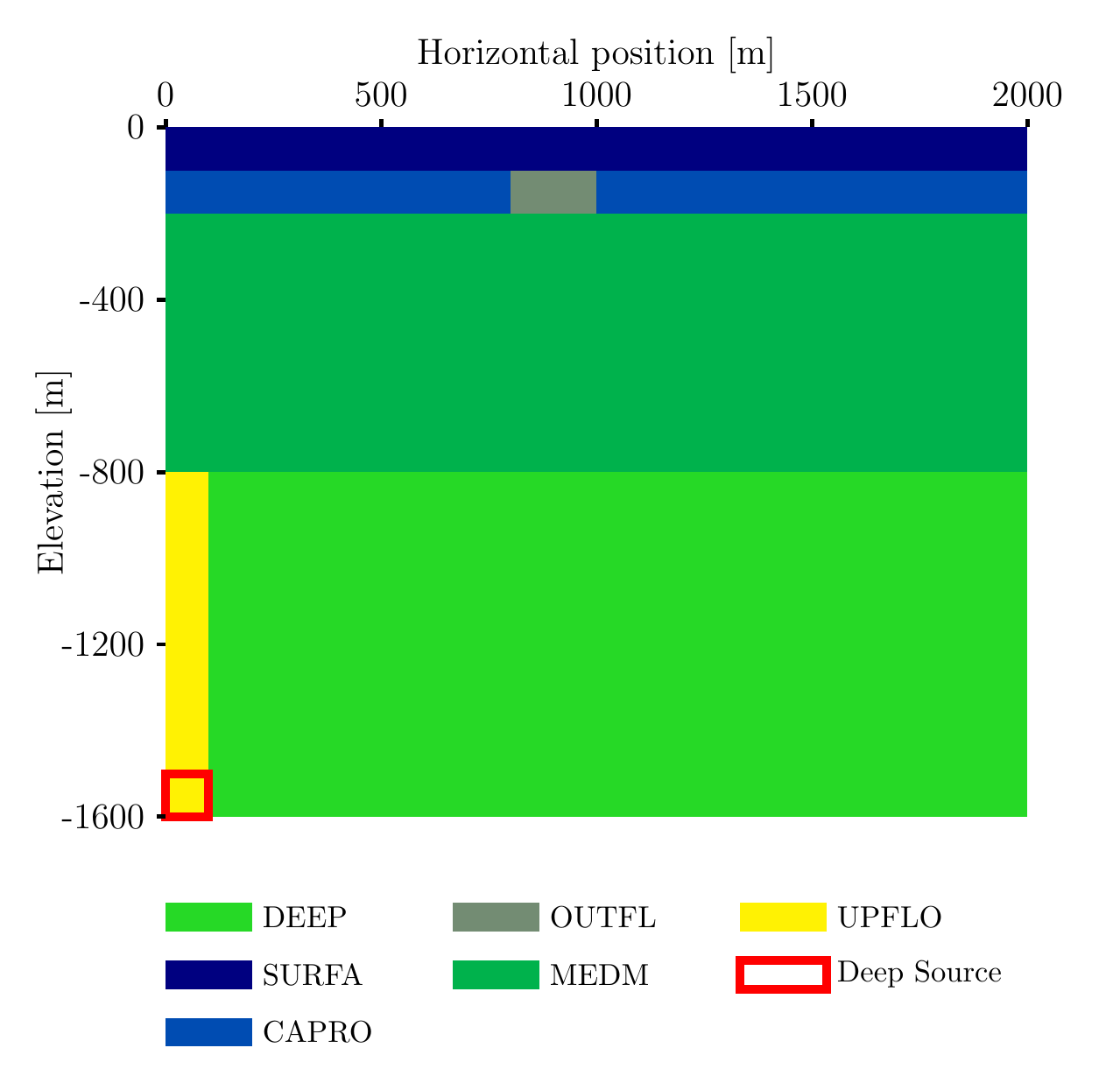}
\caption{Rectangular slice model geometry showing the rock type locations, independent of computational grid discretisation. Each rock types is represented by a different colour. The location and magnitude of the deep source in the lower left corner are assumed to be known.}\label{fig:parameter_grid}
\end{figure}

For this case study we consider a two-dimensional slice model, shown in Figure \ref{fig:parameter_grid}, based on that considered in \citeA{Bjarkason2016} and \citeA{Maclaren2016}.

The model geometry is a rectangular slice with physical dimensions of 1600 m deep and 2000 m wide. For our test problem we restricted the unknowns to a set of 12 parameters, two each for six rock type regions, where these regions are assumed known in the present work. The location and intensity of the source are also assumed known. All six rock types are assumed to have the same porosity (10\%), rock grain density (2,500 kg/m$^3$), thermal conductivity (2.5 W/(m$\cdot$K)) and specific heat (1.0 kJ/(kg$\cdot$K)). The top boundary condition consists of constant pressure of 1 atm and constant temperature of 15 $\deg$ C. The bottom boundary condition consists of a constant heat flux of 80 mW/m$^2$, except at the bottom-left corner region (see Figure \ref{fig:parameter_grid}) where 7.5 $\times 10^{-5}$ kg/(s$\cdot$m$^2$) of a 1,200 kJ/kg enthalpy fluid is used as a deep source input. The side boundaries are closed.

The (noisy) measurements consist of temperatures taken at 15 depths down each of 7 vertical wells; this gives a total of $105$ measurement points, see Figure \ref{fig:grids}. The synthetic data is corrupted by additive independent identically distributed (iid) mean zero Gaussian noise which has a standard deviation of 5 $\deg$ C.

We used two different computational discretisations, described in Section \ref{sec: AEC}.

\subsubsection{Case Study II: Kerinci Model}
\begin{figure}[H]
    \centering
    \includegraphics[width=.95\textwidth]{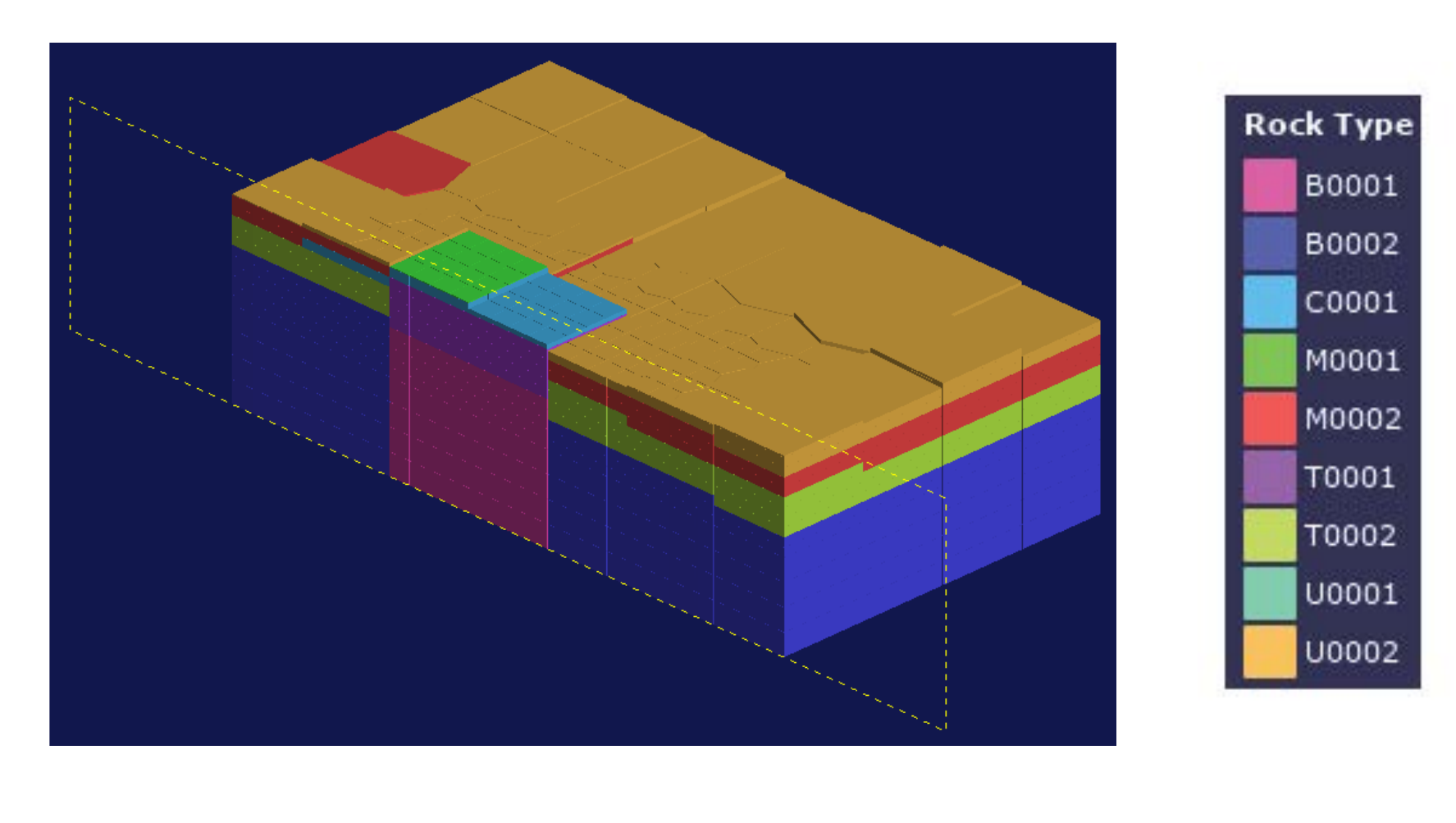}
    \caption{The Kerinci model geometry with vertical section showing the rock type locations, assumed known, based on the fine model but largely independent of computational grid discretisation. Different colours represent different rock types. The model also includes a layer of atmospheric blocks (not shown). The location and intensity of the sources are assumed to be known.}\label{fig:kerinci_geo}
\end{figure}

For this case study we consider a three-dimensional model of the Kerinci geothermal system, Sumatra, Indonesia, shown in Figure \ref{fig:kerinci_geo}. This is based on a model developed by \citeA{Prastika2016-vt}. We briefly recap the key model features here; for full details see \citeA{Prastika2016-vt}.

The model geometry has physical dimensions of 16 km by 14 km (horizontal dimensions) by 5 km (depth). Our problem has a set of 30 parameters, three each for ten rock type regions, where these regions are assumed known in the present work. One of these `rock types' corresponds to an atmospheric layer, so we have 27 key parameters of interest to estimate. All nine non-atmospheric rock types are assumed to have the same porosity (10\%), except for the rock labelled C0001 (representing pumice) which has a slightly higher porosity (12\%). The rest of the properties of the non-atmospheric rock types were uniform, assumed to have the same rock grain density (2,500 kg/m$^3$), thermal conductivity (2.5 W/(m$\cdot$K)) and specific heat (1.0 kJ/(kg$\cdot$K)). The top boundary condition consists of constant pressure of 1 bar and constant temperature of 25 $\deg$ C. Most of the bottom boundary consists of a constant heat flux of 80 mW/m$^2$, except for a small number of blocks specified as the locations of deep source input \cite<see>{Prastika2016-vt}. The total flow rate of the deep source input is 100 kg/s, split into 70 kg/s of fluid with an enthalpy of 1,400 kJ/kg and 30 kg/s of fluid with an enthalpy of 1,100 kJ/kg.

Measurements consisted of a total of 17 temperature measurements taken down 3 wells. We assume the data is corrupted by additive iid mean zero Gaussian noise with a standard deviation of 10 $\deg$ C.

We again used two different computational discretisations, described in Section \ref{sec: AEC}.

\subsection{Approximation Error Computations}\label{sec: AEC}
For each case study, we calculated the statistics of the approximation errors by using the AUTOUGH2 simulator. The same process was used in each case, though slightly different numbers of simulations were used for each case study. We outline the general process below, while indicating any differences between case studies.

\subsubsection{Calculation Steps}
In each case study, to calculate the statistics of the approximation errors, we simulated both the fine model, $\bs{f}(\kkf)$, and the coarse model, $\bs{g}(\kkc)$, 1000 times each using AUTOUGH2. These simulations were taken over the naive posterior, which was first generated by running MCMC using the coarse model and without accounting for the approximation errors 

For the slice model scenario, the naive posterior was constructed from 150,000 samples generated by MCMC, while for the Kerinci scenario we generated 90,000 samples. The statistics of the approximation errors were then calculated, as described above, by running the coarse and fine models on 1000 samples randomly selected from the full set of 150,000 (slice model) or 90,000 (Kerinci) naive posterior samples. 

For the slice model scenario the fine model geometry consisted of a square grid of 81 $\times$ 100 = 8100 blocks (including one layer of atmospheric blocks), and the coarse model consisted of a grid of 17 $\times$ 20 = 340 blocks (again including one layer of atmospheric blocks). These model grids are shown in \ref{sec: A1}.

For the Kerinci model scenario the fine model geometry consisted of 5396 blocks (including one layer of atmospheric blocks) and the coarse model consisted of 908 blocks (again including one layer of atmospheric blocks). These model grids are again shown in \ref{sec: A1}. 

In each case study we ensured consistency of measurement locations using functionality of PyTOUGH described in \cite{OSullivan2013}, which allows the same observation wells to be defined independently of grid resolution. 

\subsection{MCMC Computations}
For the slice model scenario (and both with and without incorporation of the approximation errors) 150,000 samples were computed (an ensemble of 300 {\it walkers} taking 500 samples each) after discarding an initial 30,000 {\it burn-in} samples. 

For the Kerinci scenario (both with and without incorporation of the approximation errors) 90,000 samples were computed (6 ensembles of 300 {\it walkers} taking 50 samples each) after discarding a total of 30,000 {\it burn-in} samples (5,000 for each ensemble).

All computations were carried out on a standard desktop computer with an AMD Ryzen 5 1600 3.2GHz 6-Core Processor.

\subsection{Computational Requirements of Forward Model and MCMC Simulation}
In the slice model scenario, the fine model took approximately 1-5 minutes per simulation, while the coarse model took less than half a second per simulation, typically about 0.45 seconds. Thus generating 150,000 samples using naive MCMC to construct the posterior distribution using the fine model would take around 100-500 days, whereas using the same number of samples to construct the approximation error informed posterior using the coarse model took just less than 20 hours. Only taking into account these MCMC runs, in the worst case this represents a speed-up of at least a factor of 100. 

In the Kerinci scenario, the fine model took approximately 30 seconds per simulation for well-behaved cases, but potentially several hours for less well-behaved models. The coarse model typically took about 1-10 seconds per simulation, but could take several minutes for less well-behaved cases. The run times for both of these cases were much more variable for this model than for the slice model. We generated 90,000 samples by running 6 chains in parallel (see below for more detail) and then combining these. Generating 90,000 samples to construct the posterior distribution using the fine model and naive MCMC would take at least a year, and possibly up to a decade, whereas using the same number of samples to construct the approximation error informed posterior using the coarse model and naive MCMC (again run in 6 parallel batches) took about 12 days.

More sophisticated parallelisation \cite{vrugt2009accelerating,laloy2012high}, or use of gradient information \cite{carpenter2017stan,patil2010pymc}, in the MCMC sampling algorithms could of course considerably change these timing estimates for full MCMC. Here we restrict attention to a particularly simple black-box MCMC sampler that can be easily coupled to AUTOUGH2 simulations. In general, however, we would still expect significant practical speed-ups in a range of realistic scenarios, as the BAE approach is suited to problems in which approximate pre-marginalisation can be carried out with many less samples than required for full MCMC sampling.

In addition to the above rough timing estimates, the approximation error calculations further require both a naive posterior and the model approximation error statistics to be calculated. In each case, approximately the same amount of time was required to run full MCMC for the naive case and for the approximation error informed case. The key cost for all MCMC calculations is running the coarse model; only the statistics of the particular likelihood model differ. Thus, for the slice model, approximately 20 hours was required to run full MCMC for the naive case, and another 20 hours for the approximation error informed case. For the Kerinci model, between 2-6 days would be expected to carry out naive MCMC in general; here we found it took about 2.5 days when sampling from the prior and running the naive model.

In contrast to the MCMC cases, simulations of both the coarse and accurate model are required to calculate the approximation errors statistics. For the slice model scenario we generated 1000 samples by running 200 runs of each in parallel on 5 nodes; this also took just under 20 hours. Thus the total time for inversions using (a naive version of) the approximation error approach is approximately 20 $\times$ 3 = 60 hours. The worst-case effective speed-up factor compared to naive sampling is thus at least 30 in the present work, but typically more like 50-150. 

For the Kerinci scenario we again used 200 runs of each model in parallel on 5 nodes. This took approximately 5.5 days. Thus the total time for inversions using the approximation error approach was around 2.5 + 5.5 + 6 = 14 days (2 weeks). This again gives a speed-up (compared to e.g. 1-10 years) of at least about 30.

Natural ways to further increase the speed-up of the approximation error calculations include, for example, only running an approximate, optimisation-based sampler to generate the initial naive posterior (from which only 1000 samples will be used). In our case, however, we simply ran full MCMC separately for both the naive and the approximation errors informed cases. This enabled us to give a relatively fair comparison of the results from these two models. Furthermore, the initially calculated naive posterior, or at least the second order statistics of this, could be used either to initialise the second (main) run of MCMC for our algorithm, or as a proposal distribution.

\subsection{Availability}
Our code was written in Python 2.7 using open source Python packages. It is available at:
\begin{itemize}
\item https://github.com/omaclaren/hierarchical-bae-manuscript
\end{itemize}
An archived version of this code is available at:
\begin{itemize}
    \item http://doi.org/10.5281/zenodo.3509966
\end{itemize}
Access to the AUTOUGH2/TOUGH2 \cite{Yeh2012, Pruess1999} simulator is also required; we plan to adapt our code to use the new open-source Waiwera simulator \cite{Croucher2018} when it is officially released.

The key functionality is implemented in a small library of object-oriented classes implementing the various components of the hierarchical framework. 

\section{Results and Discussion}\label{sec: results}
Here we compare a series of inversion results for both the slice model and the Kerinci model scenarios. We focus on the results that are obtained, for each scenario, when using a coarse model without accounting for approximation errors and those obtained when using a coarse model when the approximation errors are accounted for. We consider both data space (posterior predictive) and parameter space (parameter posterior) distributions.

Particular emphasis is placed on a) the feasibility of the posterior uncertainty estimates in parameter space, that is, the question of whether or not the posterior uncertainty is consistent with, i.e. supports, the true ($\log$-) permeability values, and b) the role of predictive checks with and without incorporation of the approximation errors.

\subsection{Slice Model Scenario}
Here we consider results from the slice model scenario.

\subsubsection{Posterior Predictive Checks}

\begin{figure}[H]
    \centering
    \includegraphics[width=0.99\textwidth]{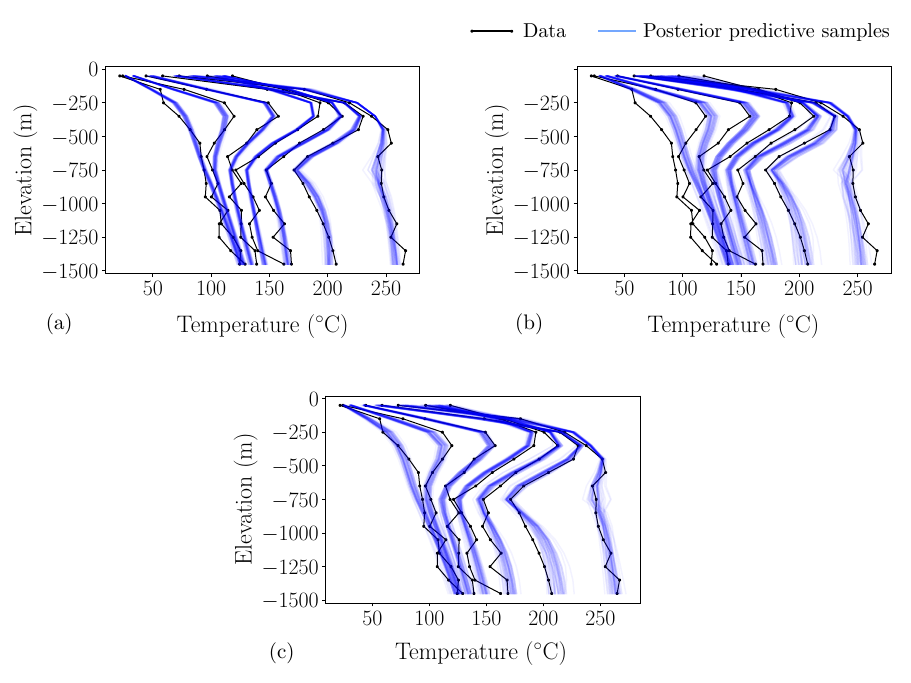}
    \caption{Posterior predictive checks for (a) the naive model without approximation error correction, (b) only the model approximation error correlations included and (c) the approximation errors included, with both the model error correlations and offset (bias) terms.} 
    \label{fig: downwelldisc}
\end{figure}

In Figure \ref{fig: downwelldisc} we show posterior predictive checks constructed by running the model on a subset of posterior samples obtained from MCMC. Realisations of the process model without measurement error are plotted in blue, while the data obtained from running the fine model and adding measurement error are shown in black. Figure \ref{fig: downwelldisc} (a) shows the posterior predictive check under the coarse model while neglecting the approximation errors. 

As can be seen in the figure, the coarse model fits the data well and the uncertainties are small. Thus this check does not flag any potential issue with naively using the coarse model. On the other hand, (b) and (c) show the predictive checks resulting from inference under the approximation error corrected model. In particular, (b) shows the results when only the covariance of the approximation errors are accounted for, while (c) shows the results when both the approximation error covariance and offset (bias) terms are included. Comparison of (b) and (c) shows that both error correlations and the bias term are important for obtaining a properly fitting model. More importantly, the difference in variation between (a) and (c) indicates that we are potentially \textit{underestimating} the uncertainties involved in naively using the coarse model for inversion. Intuitively, the low-variance of the posterior is counterbalanced by the introduction of additional bias into the parameter estimates. This is illustrated in the next subsection.

An implication of these results is that, in general, \textit{posterior predictive checks against the original data do not appear to indicate issues that arise due to inversion under a reduced-order model}. This is perhaps to be expected due to the ill-posed nature of inverse problems, i.e. pure within-sample data fit checks are not sufficient to determine whether a model is appropriate. One potential fix for this is to either carry out checks on held-out data or, in our case, against a more expensive/accurate model which effectively plays the role of held-out data. 

\subsubsection{Posterior Parameter Distributions}
\begin{figure}[H]
\centering
\includegraphics[width=1.0\textwidth]{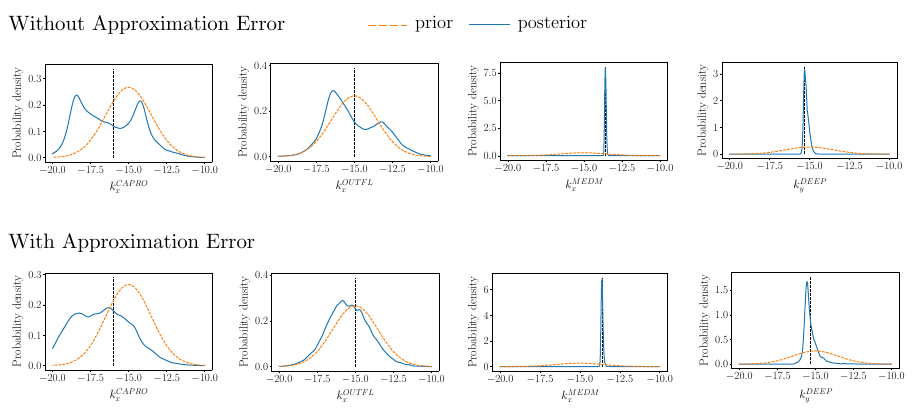}
\caption{Marginal posteriors for parameters in which the naive and approximation error informed cases largely agree. The true parameter values are indicated using a dashed black line.}\label{fig:agree-slice}
\end{figure}

Here we consider the (marginal) parameter space posterior distributions, both for the naive and the approximation error informed models. Figure \ref{fig:agree-slice} shows the marginal posteriors of the permeability for each rock type and each direction, and both with and without incorporation of the approximation errors.

\begin{figure}[H]
\centering
\includegraphics[width=1.0\textwidth]{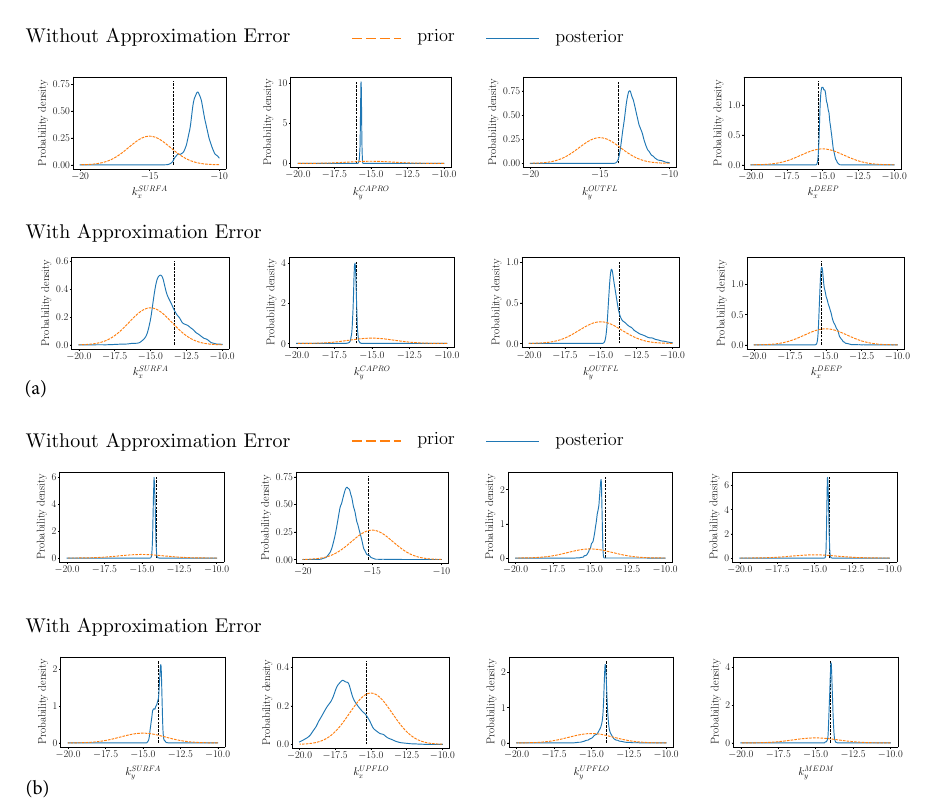}
\caption{Marginal posteriors for parameters in which the naive and approximation error informed cases tend to conflict. The true parameter values are indicated using a dashed black line. These plots are divided into two groups, labelled as (a) and (b), purely to aid visual comparison. Within each of these groupings, the posterior for each parameter is presented both above and below, with the posteriors above obtained without approximation error and those below showing the results incorporating approximation error.}\label{fig:conflict-slice}
\end{figure}

The first set of plots, in Figure \ref{fig:agree-slice}, show the parameters for which fairly consistent results were reached by both the naive and approximation error informed models. The second and third sets of plots, shown in Figure \ref{fig:conflict-slice} as two sets of $2\times4$ plots labelled by (a) and (b) for easier visual comparison, show cases where the results tend to conflict between the naive and the approximation error.

As can be seen in Figure \ref{fig:conflict-slice}, naive inversion under the coarse model often results in essentially infeasible parameter estimates, i.e. posteriors for which the truth is assigned only a low posterior probability density. On the other hand, the approximation error corrected case always assigns a high posterior probability density to the true parameters (though in \textit{some} cases this is slightly lower than the density assigned under the naive case). In reality, of course, neither model will be correct, but it is hoped that the fine-scale model is a better reflection of the truth.

Some of the parameters appear to be effectively non-identifiable, as indicated by the lack of updating when comparing the prior to posterior distributions \cite<see>[for a systematic review and discussion of measuring statistical evidence in a Bayesian setting]{Evans2015}. This lack of identifiability can also be quantified using, for example, the Kullback-Leibler divergence, however, we prefer to present comparisons graphically, following the general Bayesian data analysis philosophy of \citeA{gelman2013bayesian}. In particular $k_x^{\rm CAPRO}$ and $k_x^{\rm OUTFL}$ appear to be largely uninformed by the data. Physically this could be explained by the fact that there is very little horizontal fluid flow in the cap rock and essentially all fluid in the outflow region is in the vertical direction. On the other hand, the remaining parameters appear to be reasonably well-identifiable; some, in particular $k_y^{\rm CAPRO}$, $k_x^{\rm MEDM}$, $k_y^{\rm MEDM}$, $k_y^{\rm SURF}$ and $k_y^{\rm UPFLO}$ appear to be strongly identified. Under the naive model, however, inversion for the strongly identifiable parameters gives posteriors that \textit{appear} very well-informed but are in fact providing effectively infeasible estimates. This provides another trade-off, where parameters that are strongly informed by the data under a model will hence tend be more strongly biased towards different values when estimated under a different model.

%\begin{figure}[H]
%\centering
%\includegraphics[width=1.05\textwidth]{./%figures/to_include/%marginal_conflicting_slice_02.pdf}
%\caption{Marginal posteriors for parameters %in which the naive and approximation error %informed cases tend to conflict, the true %parameter values are indicated using a %dashed black line.}\label%{fig:conflict-slice-02}
%\end{figure}

\subsection{Kerinci Model Scenario}
Here we consider results from the Kerinci model scenario.

\subsubsection{Posterior Predictive Checks}
\begin{figure}[H]
    \centering
    \includegraphics[width=0.95\textwidth]{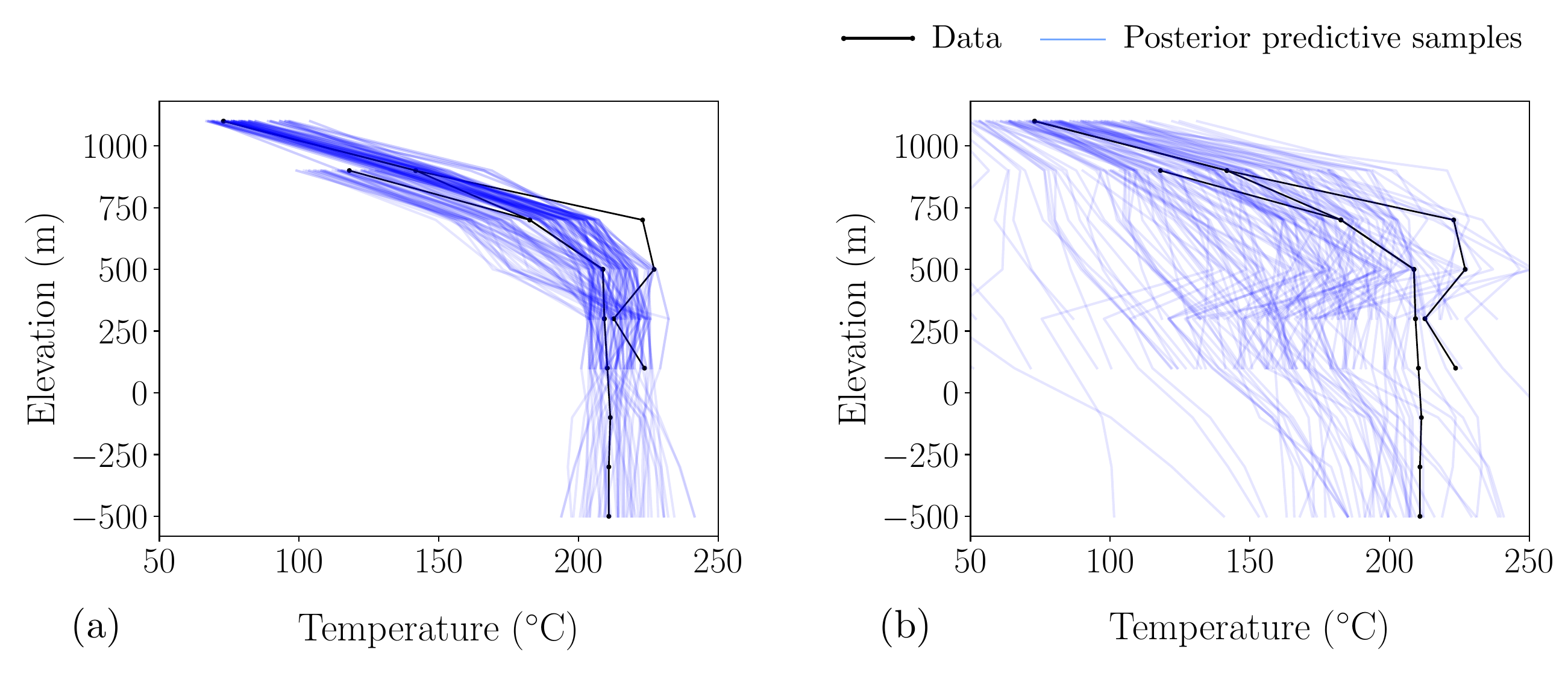}
    \caption{Posterior predictive checks for (a) the naive model without using the approximation errors, and (b) while using the approximation errors (both the model error correlations and offset/bias terms).}
    \label{fig: downwelldisc-kerinci}
\end{figure}

In Figure \ref{fig: downwelldisc-kerinci} we show posterior predictive checks constructed by running the model on a subset of posterior samples obtained from MCMC. Realisations of the process model without measurement error are plotted in blue, while the data obtained from running the fine model and adding measurement error are shown in black. Figure \ref{fig: downwelldisc-kerinci}(a) shows the posterior predictive check under the coarse model without incorporation of the approximation errors, while  Figure \ref{fig: downwelldisc-kerinci}(b) shows the results incorporating the approximation errors. 

As can be seen, in this more realistic model, and under more extreme model simplification (the discretisation is significantly reduced and simplified in the coarse model) the approximation errors can be quite large. The difference in variation between (a) and (b) in Figure \ref{fig: downwelldisc-kerinci} certaintly indicates that we are likely \textit{underestimating} the uncertainties involved in naively using the coarse model for inversion. Although the coarse model predictive check provides a tighter fit around the measured data, it also assigns much less probability density to at least one data point, so in this sense provides a worse fit to the data and in this case flags potential underfitting of the coarse model.

\subsubsection{Posterior Parameter Distributions}
\begin{figure}[H]
    \centering 
    \includegraphics[width=1.0\textwidth]{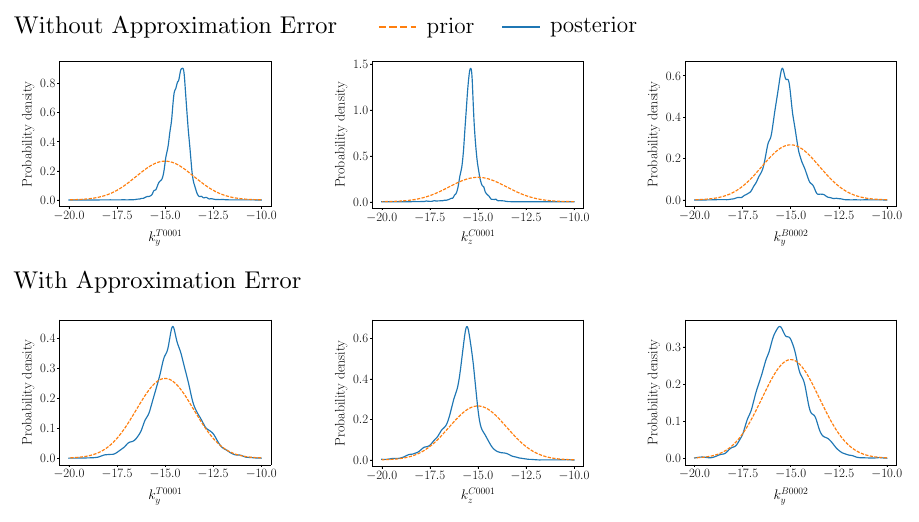}
    \caption{Marginal posteriors for the rock permeabilities in the Kerinci model, in cases where the results largely agree between the naive case and the approximation error informed case.}\label{fig:agree-kerinci}
    \end{figure}

\begin{figure}[H]
    \centering
    \includegraphics[width=1.0\textwidth]{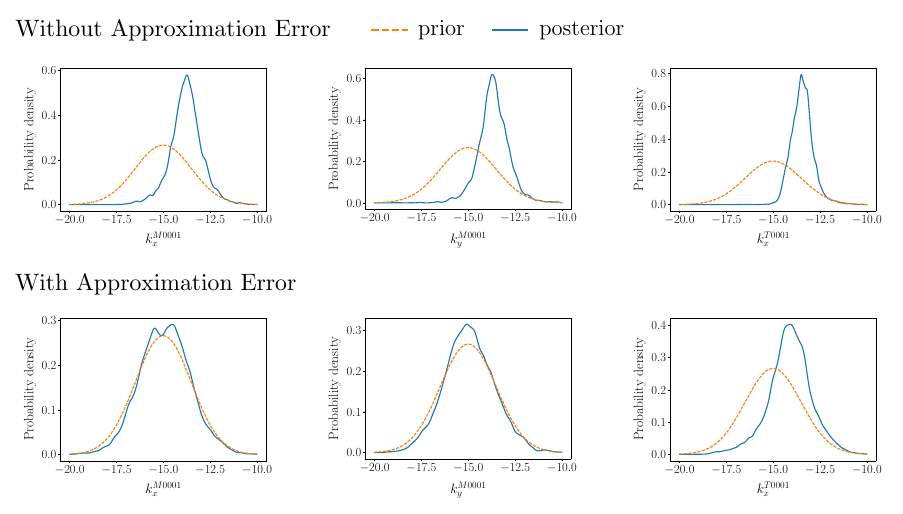}
    \caption{Marginal posteriors for the rock permeabilities in the Kerinci model, in cases where the results tend to conflict between the naive case and the approximation error informed case.}\label{fig:conflict-kerinci}
    \end{figure}

Here we consider the (marginal) parameter space posterior distributions, both for the naive and the approximation error informed models, see Figures \ref{fig:agree-kerinci} and \ref{fig:conflict-kerinci}. For brevity we include a representative selection here; the remaining distributions as well as full corner plots \cite{Foreman-Mackey2016} are given in the supplementary information (SI). The same basic patterns observed in the plots shown here can also be seen in the plots in the SI. 

The first set of plots, in Figure \ref{fig:agree-kerinci}, show the parameters for which fairly consistent results were reached by both the naive and approximation error informed models. The second set of plots, shown in Figure \ref{fig:conflict-kerinci}, shows cases where the results tend to conflict between the naive and the approximation error informed case. Here the true parameters are unknown and hence not shown.

\subsection{Additional Comments}
As we have noted above, standard MCMC sampling is much more computationally feasible for these geothermal inverse problems when using coarser models as opposed to finer, more accurate models. Importantly, however, we see that just naively using a coarse model without accounting for approximation errors tends to give overconfident and biased posteriors, for which the known true parameters can lie outside of the bulk of the support. On the other hand, taking into account the approximation errors leads to known true parameters lying inside the bulk of the support in all cases considered here. Both methods require effectively the same amount of computation time, though the BAE approach requires some additional initial computation to construct the model error statistics. This additional computational effort is the price paid to avoid misleading estimates and is still significantly less than attempting MCMC using the fine model. 

In this paper we have only considered the use of relatively naive MCMC sampling to estimate the posterior density for the permeabilities, based on an approximation error informed coarse model. More sophisticated MCMC algorithms, e.g. those utilising parallelisation \cite{vrugt2009accelerating,laloy2012high} and/or derivative information \cite{carpenter2017stan,patil2010pymc} would be expected to speed up the sampling significantly. In some settings, however, these more sophisticated forms of MCMC may still be computationally infeasible even using only the coarse model (with or without approximation errors included). In this case, the posterior approximation errors can still be constructed without MCMC, as long as some alternative method is available for drawing the (smaller) set of required samples from the naive posterior. For example, here we only required 1000 samples from the naive posterior, compared to the 150,000 or 90,000 used for full MCMC runs. This would then enable the use of a coarse model which accounts for approximation errors alongside alternative sampling and/or optimisation-based approaches. 

\section{Conclusions}
We have demonstrated how to carry out simple yet computationally feasible parameter estimation and uncertainty quantification for geothermal simulation models by using a coarser, or cheaper, model in place of a finer, or more expensive, model. Our approach was to construct an approximation to the posterior Bayesian model approximation error and incorporate this into a hierarchical Bayesian framework. The hierarchical Bayesian perspective provides a flexible and intuitive setting for specifying assumptions on different model components and their combinations. In this view, approximations and modelling assumptions are directly incorporated into the framework by replacing joint distributions by factorisations in terms of simpler conditional and/or marginal distributions. 

Our approach requires two simple initial computational steps in order to correct for the bias and/or overconfidence that would normally be introduced by directly using the coarse model in place of the finer model. These two steps then enable standard, out-of-the-box MCMC to be used to sample the parameter posterior using the coarse model. We demonstrated our approach can achieve significant computational speed-ups on both synthetic and real world geothermal test problems.  

Our approach consists of three relatively simple steps overall and should be more accessible to general practitioners than having to manually implement more complex sampling schemes. Furthermore, the methods developed here should be generally applicable to related inverse problems such as, for example, those appearing in petroleum reservoir engineering and groundwater management.

%%%%%%%%%%%%%%%%%%%%%%%%%%%%%%%%
%% Optional Appendix goes here
%
% The \appendix command resets counters and redefines section heads
%
% After typing \appendix
%
%\section{Here Is Appendix Title}
% will show
% A: Here Is Appendix Title
%
\appendix
%\section{Here is a sample appendix}
\section{Mapping Between Fine and Coarse Grids}\label{sec: A1}
\begin{figure}[h]
\centering
\includegraphics[width=1.0\textwidth]{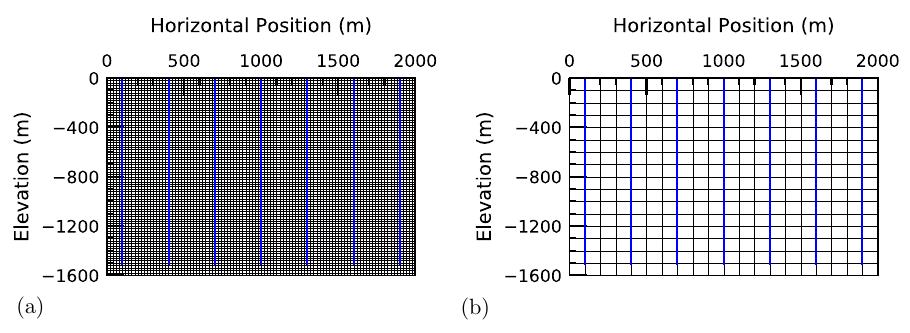}
\caption{Computational grids used to simulate the geothermal slice system. The fine model geometry (a) consisted of a square grid of 81 $\times$ 100 = 8100 blocks (including one layer of atmospheric blocks), and the coarse model (b) consisted of a grid of 17 $\times$ 20 = 340 blocks (again including one layer of atmospheric blocks). Observation wells are shown as blue vertical lines.}\label{fig:grids}
\end{figure}

To facilitate computation of the process model error, and following \cite{Kaipio2013}, in this study we have been implicitly assuming that the difference between the fine and coarse models can be approximated as
\begin{equation}
\eps = \yp - \g(\kkc) = \f(\kkf) - \g(\kkc) \approx \f(\kkc) - \g(\kkc) = \f(\kk) - \g(\kk).
\end{equation}
Here $\f(\kkc)$ is a slight abuse of notation and still in fact represents the fine model evaluated on a fine parameter grid; however the parameters are fixed to be homogeneous within a given rock type, matching the values in the corresponding rock types on the coarser grid. Thus the parameter vectors have the same effective dimension (and values), equal to that of the coarse grid, and thus are in 1-1 correspondence. This is made clearer by comparing Figure \ref{fig:grids} below to Figure \ref{fig:parameter_grid} introduced earlier: each mesh in Figure \ref{fig:grids} represents a different discretisation of the \textit{same} underlying parameter grid given in Figure \ref{fig:parameter_grid}. This assumption means we can compute the approximation error by sampling the coarse parameters directly rather than the (larger-dimensional) fine parameters. Implicitly, however, this is neglecting some of the approximation error that would be induced by sampling over all fine parameter sets compatible with the given coarse parameter set. This assumption can be checked/removed to the extent that computational resources allow computing the error over the fine grid \cite{Kaipio2013}. Either way, the \textit{coarse} grid parameters are the ultimate targets of inference, and by using the more conservative `enhanced' (or `composite') error model based on the marginal error distribution we can hope to account for some of this additional uncertainty indirectly. 

\begin{figure}[h]
    \centering
    \includegraphics[width=1.0\textwidth]{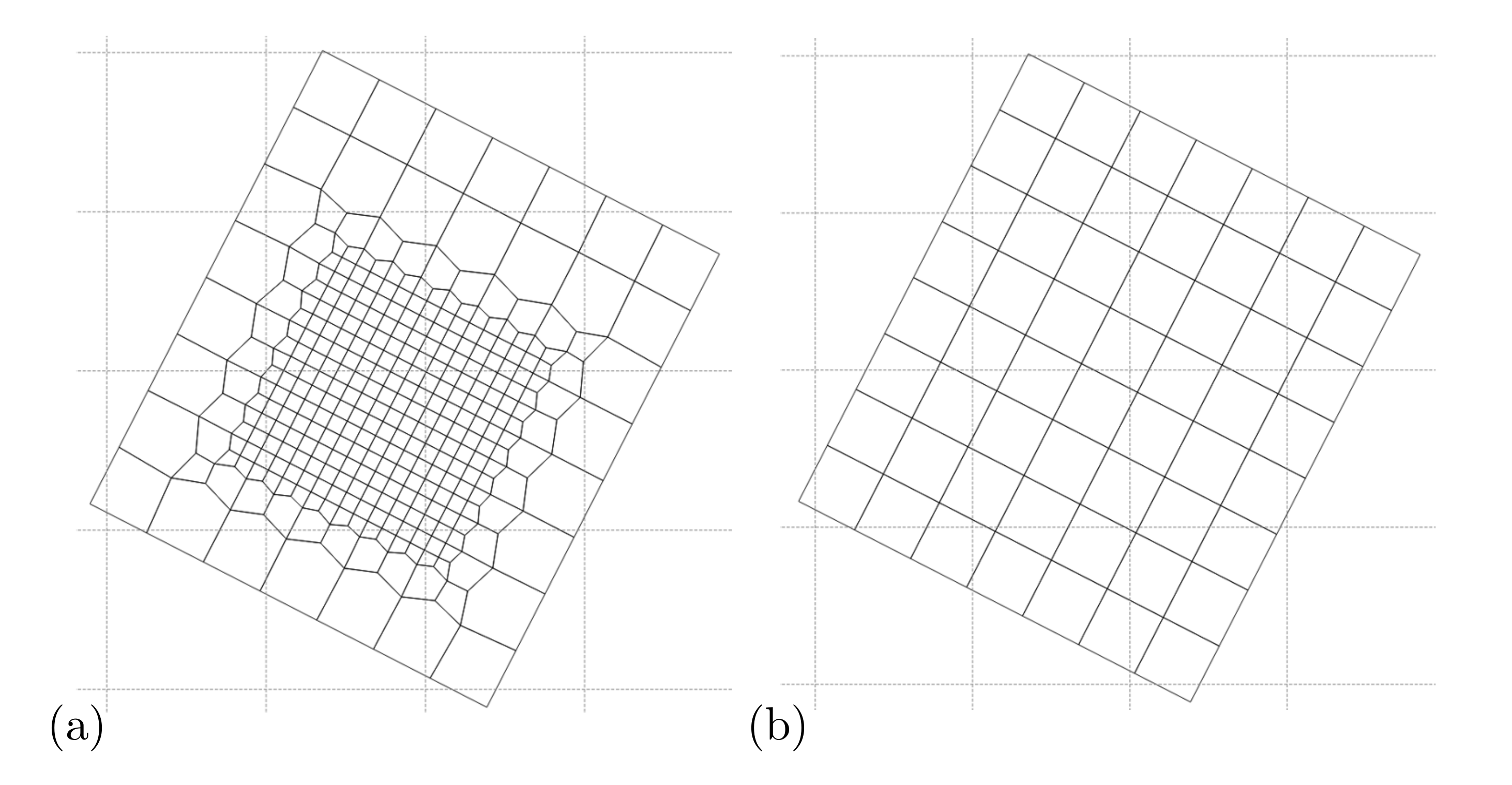}
    \caption{Computational grids used to simulate the Kerinci geothermal system, showing only the top-down view for illustrative purposes. The fine model geometry (a) consisted of a 3D system of 5396 blocks (including one layer of atmospheric blocks), and the coarse model (b) consisted of a 3D system of 908 blocks (again including one layer of atmospheric blocks).}\label{fig:kerinci_grids} 
    \end{figure}

The fine and coarse Kerinci models were related in the same manner, with only the mesh discretisation varying. A top view of the two meshes in shown in Figure \ref{fig:kerinci_grids}.

\section{Geometric View of BAE}\label{sec: A2}
\begin{figure}[H]
\centering
\includegraphics[width=1.0\textwidth]{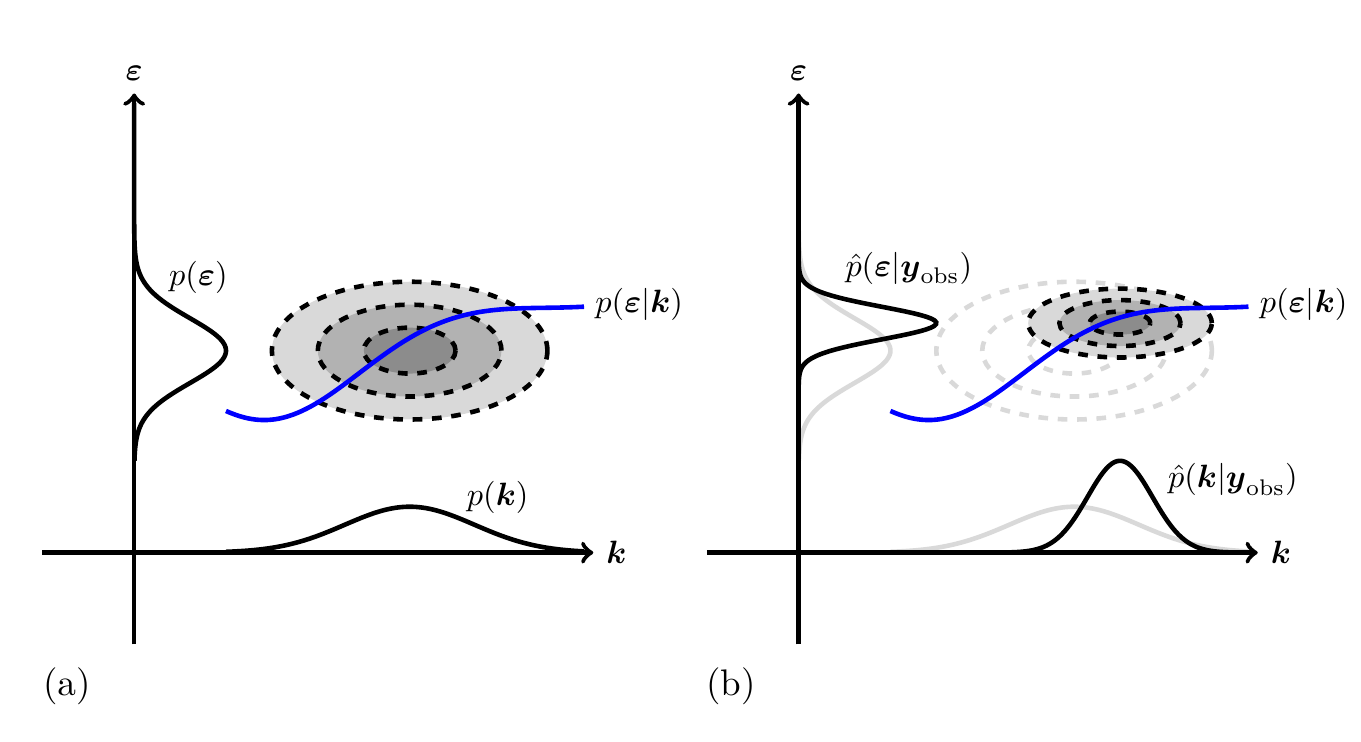}
\caption{Geometric interpretation of the enhanced/composite Bayesian model approximation error approach in both (a) the usual prior-based case and (b) our posterior-based approach. In both cases we essentially aim to conservatively cover the deterministic functional relationship $p(\eps | \kk)$, or the associated degenerate joint distribution $p(\eps | \kk)p(\kk)$, by a probability distribution based on marginal distributions. In the posterior case, however, we restrict attention to estimating the error by sampling over the support of the naive posterior.}\label{fig:composite_error}
\end{figure}

In Figure \ref{fig:composite_error} below we give a geometric picture of both the standard prior-based and our posterior-based composite (enhanced) error model approach. In both cases we essentially aim to conservatively cover the deterministic functional relationship $p(\eps | \kk)$, or the associated degenerate joint distribution $p(\eps | \kk)p(\kk)$, by a probability distribution based on marginal distributions. In the posterior case, however, we restrict attention to estimating the error by sampling over the support of the naive posterior. As can be seen in the figure, the accuracy of this procedure depends on, for example, how well the naive posterior approximates the true posterior. Alternatively, if the error is approximately independent of the parameter, hence giving a horizontal line for $p(\eps | \kk)$, then both the prior and posterior error distributions would give the same delta distribution for the error, regardless of how well the naive posterior approximates the true posterior. Thus, intuitively, the procedure would be expected to be most reasonable when a) the naive posterior approximates the true posterior reasonably well and/or b) when the model error does not depend strongly on the parameter. This latter condition is already a condition for the usual enhanced/composite error model approach to providing a reasonable approximation, and so switching to the posterior composite error model is at least consistent with this assumption.

\section{An illustrative Example}\label{sec: A3}
Here we consider a simple curve-fitting problem to provide some further intuition for our method, and to provide an example of how the method works with correlated measurement errors.  We take the accurate model between parameters and observations to be given by $y_i=\sum_{\ell=1}^n k_\ell t_i^\ell=f_i(\bs{k})$, for $i=1,2,\dots,m$ i.e. an $n-$th order polynomial measured at points $t_1,t_2,\dots t_m$, and take the coarse model to be given by $g_i(\bs{k})=\sum_{\ell=1}^p k_\ell t_i^\ell$, for some $p<n$.

We assume a Gaussian prior and zero mean Gaussian additive noise, i.e., $\bs{k}\sim\mathcal{N}(\bs{k}_*,\bs{\Gamma}_k)$ and $\bs{e}\sim\mathcal{N}(\bs{0},\bs{\Gamma}_e)$. The fact that both forward models are linear (in $\bs{k}$), and that both the prior and the noise distribution are Gaussian means that the resulting posterior is also Gaussian. Furthermore, no (MCMC) sampling is required. Linearity of both the models allows us to write 
\begin{align}
f(\bs{k})=\bs{F}\bs{k}\text{\quad and \quad} g(\bs{k})=\bs{G}\bs{k}.
\end{align}
In this case we have $\bs{G}=\bs{F}\bs{P}$ where $\bs{P}\in\mathbb{R}^{n\times n}$ is the diagonal orthogonal projection matrix, $P_{\ell \ell}=1$ for $1\leq \ell\leq p$ and zero otherwise. This results in the standard BAE composite posterior\footnote{In this simple case it is in fact possible to use the standard BAE approach as model failures are not an issue.} coinciding with the posterior computed using our proposed approach.

We compare a) the naive approach, i.e., ignoring the model approximation errors all together, b) the posterior densities computed using our proposed approach and c) the true posterior, calculated using the accurate forward model, $f$. 

In all cases the same prior is used, while the associated likelihoods are modified. The naive posterior, the posterior based on the proposed approach, and the true posterior are given by 
\begin{align}
\hat{p}(\bs{k}|\bs{y}_{\rm obs})&\propto\exp\left(-\frac{1}{2}(\bs{k}-\hat{\bs{k}}_{\rm MAP})^T\hat{\bs{\Gamma}}^{-1}(\bs{k}-\hat{\bs{k}}_{\rm MAP})\right),\nonumber\\
p(\bs{k}|\bs{y}_{\rm obs})&\propto\exp\left(-\frac{1}{2}(\bs{k}-\bs{k}_{\rm MAP})^T\bs{\Gamma}^{-1}(\bs{k}-\bs{k}_{\rm MAP})\right),\nonumber\\
\mathring{p}(\bs{k}|\bs{y}_{\rm obs})&\propto\exp\left(-\frac{1}{2}(\bs{k}-\mathring{\bs{k}}_{\rm MAP})^T\mathring{\bs{\Gamma}}^{-1}(\bs{k}-\mathring{\bs{k}}_{\rm MAP})\right),
\end{align}
respectively. The associated MAP estimates are given by
\begin{align}
\hat{\bs{k}}_{\rm MAP}:&=\arg\min_{\bs{k}\in\mathbb{R}^n} \left\{\norm{\bs{L}_e(\bs{y}_{\rm obs}-\bs{G}{\bs{k}})}^2+\norm{\bs{L}_{{k}}({\bs{k}}-{\bs{k}}_*)}^2\right\},\nonumber\\
\bs{k}_{\rm MAP}:&=\arg\min_{\bs{k}\in\mathbb{R}^n}\left\{\norm{\bs{L}_\nu(\bs{y}_{\rm obs}-\bs{G}{\bs{k}}-\bs{\nu}_*)}^2+\norm{\bs{L}_{{k}}({\bs{k}}-{\bs{k}}_*)}^2\right\},\nonumber\\
\mathring{\bs{k}}_{\rm MAP}:&=\arg\min_{\bs{k}\in\mathbb{R}^n} \left\{\norm{\bs{L}_e(\bs{y}_{\rm obs}-\bs{F}{\bs{k}})}^2+\norm{\bs{L}_{{k}}({\bs{k}}-{\bs{k}}_*)}^2\right\},
\end{align}
and the posterior covariance matrices
\begin{align}
\hat{\bs{\Gamma}}=(\bs{G}^T\bs{\Gamma}_e^{-1}\bs{G}+\bs{\Gamma}_k^{-1})^{-1},\quad\bs{\Gamma}=(\bs{G}^T\bs{\Gamma}_\nu^{-1}\bs{G}+\bs{\Gamma}_k^{-1})^{-1},\quad\mathring{\bs{\Gamma}}=(\bs{F}^T\bs{\Gamma}_e^{-1}\bs{F}+\bs{\Gamma}_k^{-1})^{-1},
\end{align}
where $\bs{L}_e^T\bs{L}_e=\bs{\Gamma}_e^{-1}$, $\bs{L}_k^T\bs{L}_k=\bs{\Gamma}_k^{-1}$,  $\bs{L}_\nu^T\bs{L}_\nu=\bs{\Gamma}_\nu^{-1}$, with $\bs{\Gamma}_\nu=(\bs{F}-\bs{G})\hat{\bs{\Gamma}}(\bs{F}-\bs{G})^T$, and $\bs{\nu}_*=(\bs{F}-\bs{G})\hat{\bs{k}}_{\rm MAP}$.

Inline with the geothermal examples, we take the prior covariance matrix for $\bs{k}$ to be diagonal, i.e. $\bs{\Gamma}_k=\delta_k^2\bs{I}_{n}$, with $\bs{I}_{n}$ denoting the $n\times n$ identity matrix. This choice of prior, along with the fact that the simplified model is of the form $\bs{G}=\bs{F}\bs{P}$, in fact results in the posterior of our proposed method being identical to the true posterior for the first $p$ parameters, i.e., $k_1,k_2,\dots,k_p$. 

\begin{figure}[h!]
    \centering
    \includegraphics[width=0.9\linewidth]{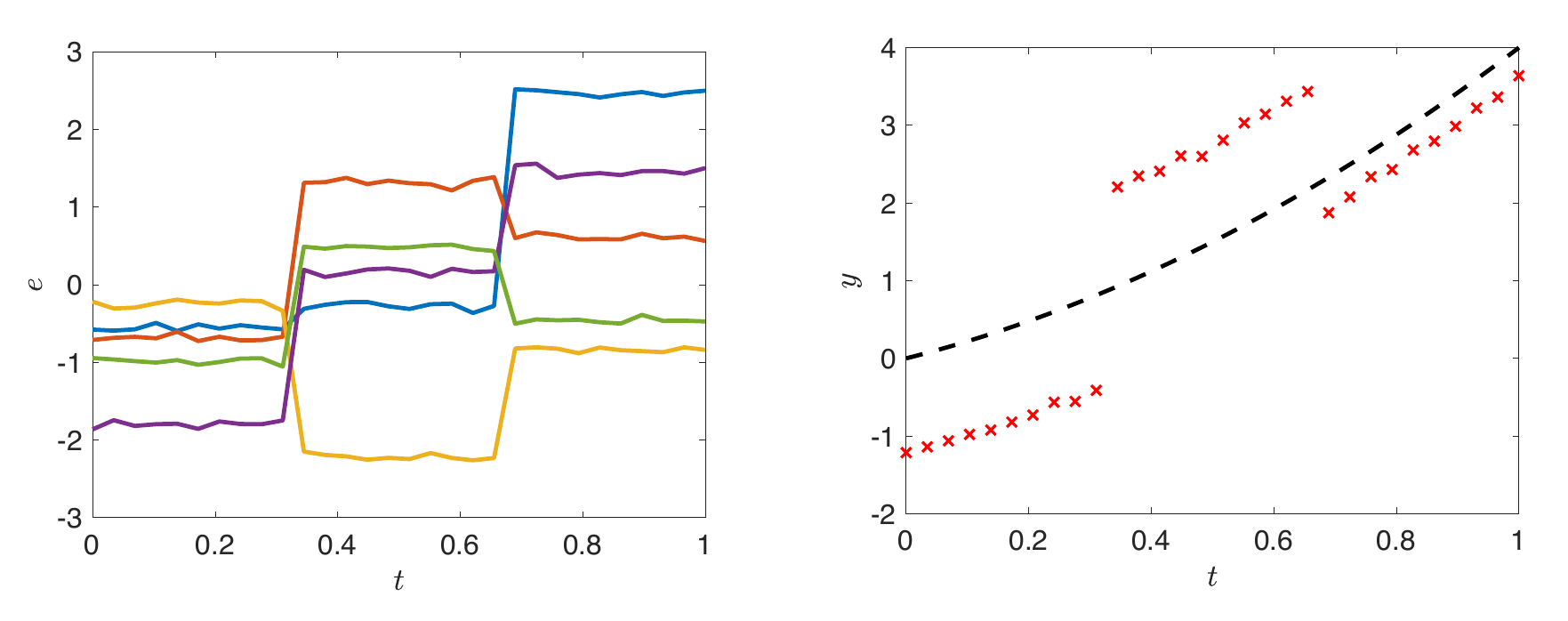}
    \caption{Five draws from the correlated noise prior distribution (left) and the data used for this synthetic example (red crosses) along with the true underlying model (black dashed line). }\label{pic: NoiseData}
    \end{figure}

To demonstrate how the method works with \textit{correlated} measurement noise, we take the additive noise to be of the (multi-level) form
\begin{align}
\bs{\Gamma}_e=\delta_e^2((1-c)\bs{D}+c\bs{I}_m),
\end{align}
where $\bs{D}$ is a block diagonal matrix of the form
\begin{align}
\bs{D}=\begin{bmatrix}
\mathds{1} & \mathbb{O} & \mathbb{O} &\dots & \mathbb{O}\\
 \mathbb{O} & \mathds{1}& \mathbb{O} & \dots & \mathbb{O}\\
 \vdots & & \ddots  & & \vdots\\
  \mathbb{O} &  \dots &\mathbb{O} & \mathds{1} & \mathbb{O}\\
    \mathbb{O} &  \dots & \mathbb{O} &\mathbb{O} & \mathds{1}\\
\end{bmatrix}\in\mathbb{R}^{m\times m},
\end{align}
with $\mathds{1}$ and $\mathbb{O}$ used to denote the square matrices of all ones and all zeros respectively. Several draws of correlated errors are shown in Figure \ref{pic: NoiseData}.

For this example we specify the number of measurements as $m=30$, with measurement points equally spaced between $t_1=0$ and $t_m=1$. For ease of visualisation we take the accurate model to be a quadratic, while the coarse model is linear, i.e., $n=2$ and $p=1$. The prior mean is $k_*=[1,1]^T$, and we take $\delta_k=1$. Finally, the correlated noise distribution is set by taking $\delta_e=1.2$, $c=0.001$, and setting the block diagonal matrix $\bs{D}$ to have three $10\times 10$ diagonal blocks, this corresponds to a noise level of $30\%$ of the maximum of the noiseless synthetic measurements, i.e., $\delta_e=\frac{30}{100}\times \max\left\{\bs{F}\bs{k}_{\rm true}\right\}$, see Figure \ref{pic: NoiseData}. Also shown in Figure \ref{pic: NoiseData} is the data for this example.

\begin{figure}[h!]
    \centering
    \includegraphics[width=.9\linewidth]{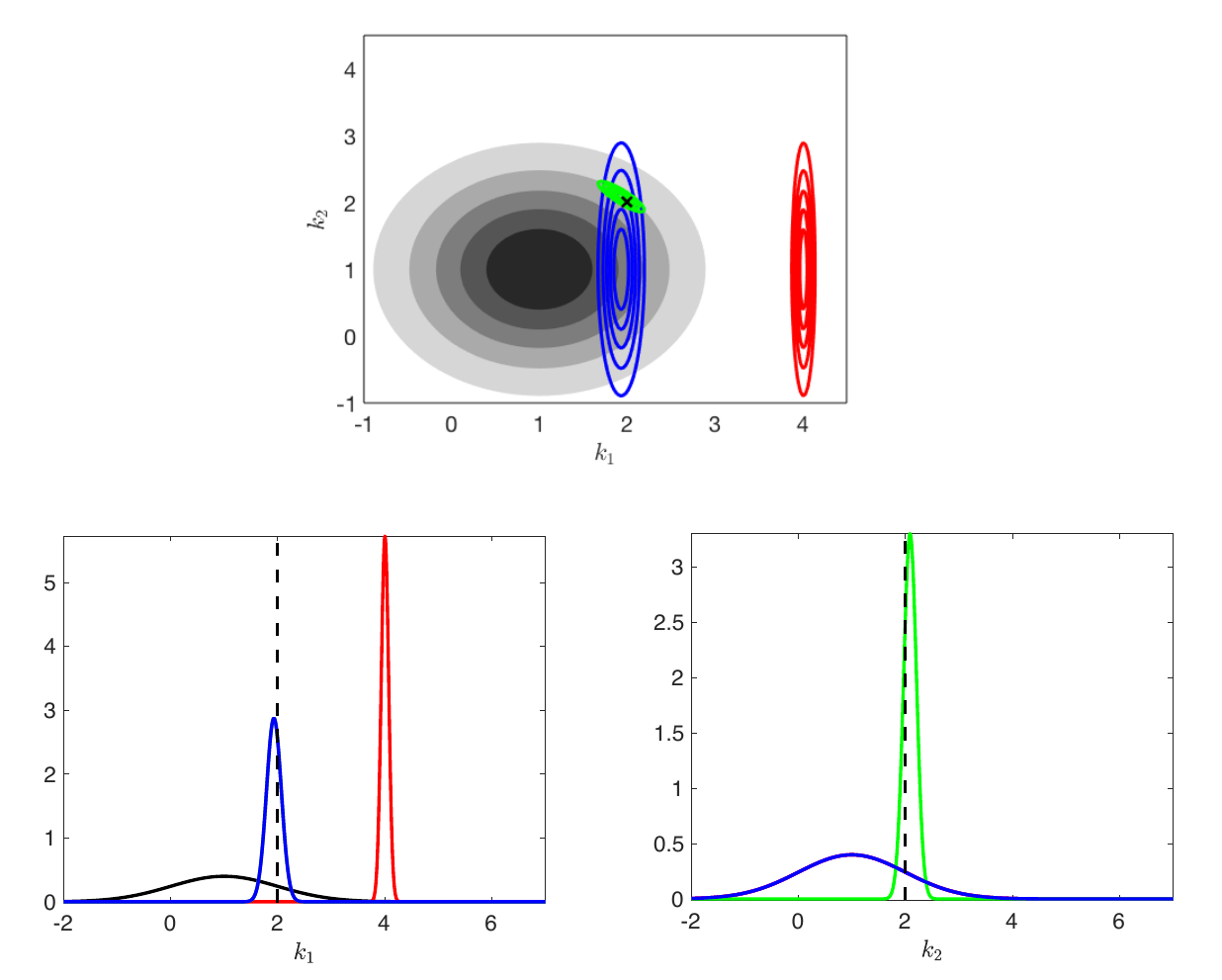}
    \caption{Joint distributions (top) and marginal distributions for $k_1$ (left bottom) and $k_2$ (right bottom). The prior is shown using either grayscale or a solid black line, the true posterior is shown in green, the naive posterior is shown in red, the posterior found using the proposed approach is shown in blue, and the true values are identified with either black cross or a dashed black line. Note that in the marginal plot for $k_1$ the true posterior is identical to marginal posterior found using the proposed method, furthermore, in the marginal plot for $k_2$ both posterior marginals using the simpler model are equal to the prior marginal.}\label{pic: Post}
    \end{figure}

\begin{figure}[h!]
        \centering
        \includegraphics[width=0.9\linewidth]{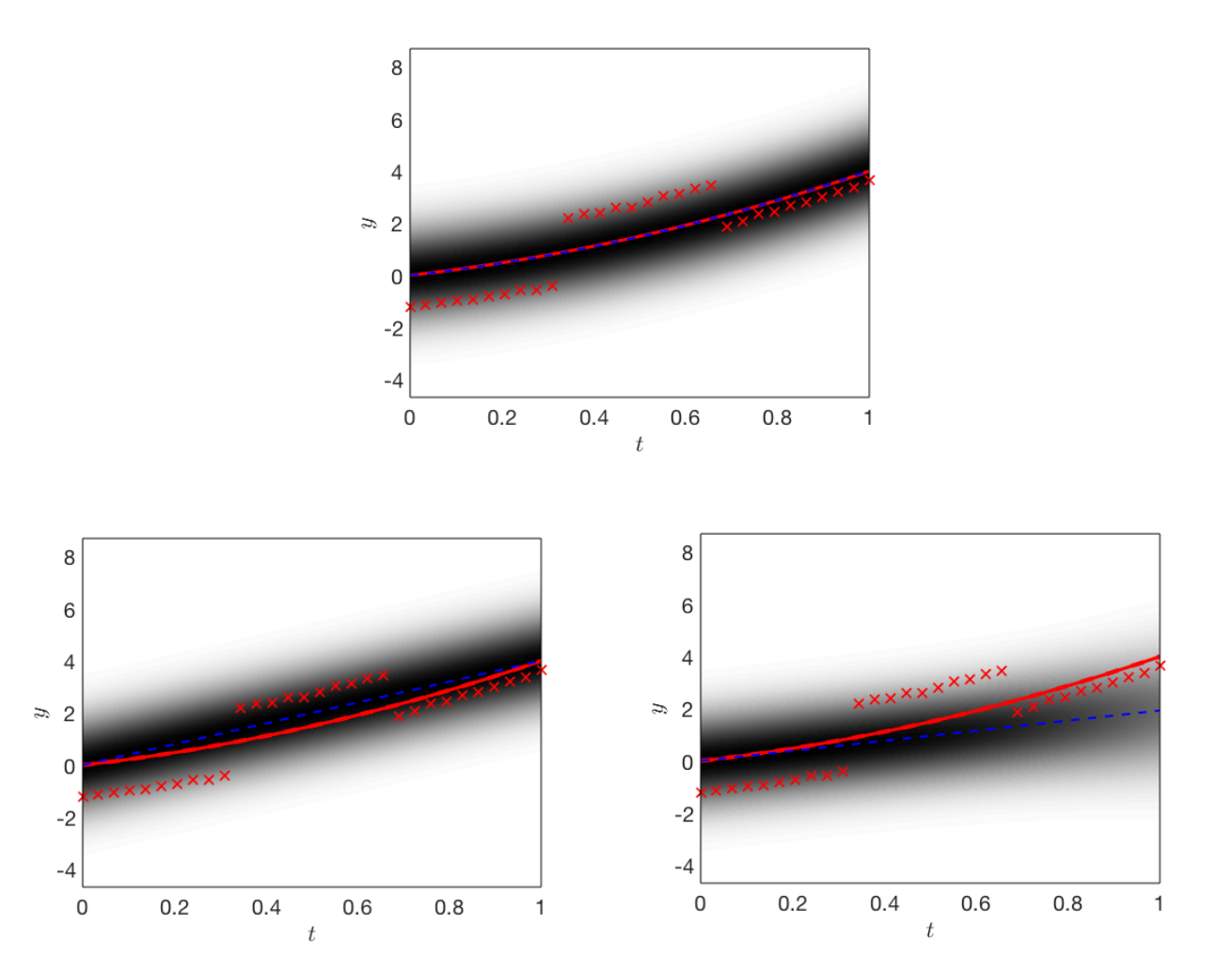}
        \caption{Posterior predictive plots using the true posterior (top), the naive posterior (left bottom), and the posterior composite error model posterior (right bottom). In all plots the data is indicated with red crosses, the true underlying model is shown with the red solid line, the predictive mean is shown with the blue dashed line, while the uncertainty in the prediction intervals is indicated using grayscale, with higher probability density indicated by darker shading. }\label{pic: Pred}
        \end{figure}

The resulting marginal and joint posteriors for $k_1$ and $k_2$ using each of the methods are shown in Figure \ref{pic: Post}, while the posterior predictive plots are shown in Figure \ref{pic: Pred}. It is clear that using the naive posterior (i.e., neglecting the approximation errors) can lead to an infeasible posterior, in the sense that the true values for $\bs{k}$ have almost vanishing posterior probability. On the other hand, in this example, using the proposed posterior composite error model leads to a feasible posterior with a more representative MAP estimate.

\acknowledgments
The authors appreciate the contribution of the NZ Ministry of Business, Innovation and Employment for funding parts of this work through the grant: C05X1306 “Geothermal Supermodels”. The authors would also like to thank Jari Kaipio for helpful discussions about Bayesian approximation error methods, Joris Popineau for visualisations of the Kerinci model, Ryan Tonkin for useful discussions on geothermal modelling, and the three reviewers for feedback that significantly improved this manuscript. Our code is available from:\\\noindent https://github.com/omaclaren/hierarchical-bae-manuscript\\\noindent and is archived at:\\\noindent http://doi.org/10.5281/zenodo.3509966

%% ------------------------------------------------------------------------ %%
%% References and Citations

%%%%%%%%%%%%%%%%%%%%%%%%%%%%%%%%%%%%%%%%%%%%%%%
% BibTeX is preferred:
%
% \bibliography{<name of your .bib file>}
\bibliography{Master}
%
% don't specify bibliographystyle
%%%%%%%%%%%%%%%%%%%%%%%%%%%%%%%%%%%%%%%%%%%%%%%

% Please use ONLY \citeA and \cite for reference citations.
% DO NOT use other cite commands (e.g., \cite, \citeyear, \nocite, \citealp, etc.).
%% Example \citeA and \cite:
%  ...as shown by \citeA{Boug10}, \citeA{Buiz07}, \citeA{Fra10},
%  \citeA{Ghel00}, and \citeA{Leit74}.

%  ...as shown by \cite{Boug10}, \cite{Buiz07}, \cite{Fra10},
%  \cite{Ghel00, Leit74}.

%  ...has been shown \cite [e.g.,][]{Boug10,Buiz07,Fra10}.

\end{document}